\input lanlmac
\def\href#1#2{{#2}}
\def\hhref#1{{#1}}
\input epsf.tex

\overfullrule=0mm

\newcount\figno
\figno=0
\def\fig#1#2#3{
\par\begingroup\parindent=0pt\leftskip=1cm\rightskip=1cm\parindent=0pt
\baselineskip=11pt
\global\advance\figno by 1
\midinsert
\epsfxsize=#3
\centerline{\epsfbox{#2}}
\vskip 12pt
{\bf Fig.\ \the\figno:} #1\par
\endinsert\endgroup\par
}
\def\figlabel#1{\xdef#1{\the\figno}}
\def\encadremath#1{\vbox{\hrule\hbox{\vrule\kern8pt\vbox{\kern8pt
\hbox{$\displaystyle #1$}\kern8pt}
\kern8pt\vrule}\hrule}}


\def\IR{\relax{\rm I\kern-.18em R}}
\font\cmss=cmss10 \font\cmsss=cmss10 at 7pt

\font\cmss=cmss10 \font\cmsss=cmss10 at 7pt
\def\IZ{\relax\ifmmode\mathchoice
{\hbox{\cmss Z\kern-.4em Z}}{\hbox{\cmss Z\kern-.4em Z}}
{\lower.9pt\hbox{\cmsss Z\kern-.4em Z}}
{\lower1.2pt\hbox{\cmsss Z\kern-.4em Z}}\else{\cmss Z\kern-.4em Z}\fi}
\def\IN{\relax{\rm I\kern-.18em N}}
\def\b{\circ}
\def\n{\bullet}

\def\gbbbb{\Gamma_4^{\hbox{$\scriptstyle \b \b$}\kern -8.2pt
\raise -4pt \hbox{$\scriptstyle \b \b$}}}
\def\gnnnn{\Gamma_4^{\hbox{$\scriptstyle \n \n$}\kern -8.2pt  
\raise -4pt \hbox{$\scriptstyle \n \n$}}}
\def\gnnnnnn{\Gamma_6^{\hbox{$\scriptstyle \n \n \n$}\kern -12.3pt
\raise -4pt \hbox{$\scriptstyle \n \n \n$}}}
\def\gbbbbbb{\Gamma_6^{\hbox{$\scriptstyle \b \b \b$}\kern -12.3pt
\raise -4pt \hbox{$\scriptstyle \b \b \b$}}}
\def\gbbbbc{\Gamma_{4\, c}^{\hbox{$\scriptstyle \b \b$}\kern -8.2pt
\raise -4pt \hbox{$\scriptstyle \b \b$}}}
\def\gnnnnc{\Gamma_{4\, c}^{\hbox{$\scriptstyle \n \n$}\kern -8.2pt
\raise -4pt \hbox{$\scriptstyle \n \n$}}}
\def\Rbud#1{{\cal R}_{#1}^{-\kern-1.5pt\blacktriangleright}}
\def\Rleaf#1{{\cal R}_{#1}^{-\kern-1.5pt\vartriangleright}}
\def\Rbudb#1{{\cal R}_{#1}^{\circ\kern-1.5pt-\kern-1.5pt\blacktriangleright}}
\def\Rleafb#1{{\cal R}_{#1}^{\circ\kern-1.5pt-\kern-1.5pt\vartriangleright}}
\def\Rbudn#1{{\cal R}_{#1}^{\bullet\kern-1.5pt-\kern-1.5pt\blacktriangleright}}
\def\Rleafn#1{{\cal R}_{#1}^{\bullet\kern-1.5pt-\kern-1.5pt\vartriangleright}}
\def\Wleaf#1{{\cal W}_{#1}^{-\kern-1.5pt\vartriangleright}}
\def\Cleaf{{\cal C}^{-\kern-1.5pt\vartriangleright}}
\def\Cbud{{\cal C}^{-\kern-1.5pt\blacktriangleright}}
\def\Crleaf{{\cal C}^{-\kern-1.5pt\circledR}}


\magnification=\magstep1
\baselineskip=12pt
\hsize=6.3truein
\vsize=8.7truein
 at 8truept
 at 8truept
 at 10truept

\font\bigrm=cmr12 at 14pt
\centerline{\bigrm Distance statistics in quadrangulations with no 
multiple edges}
\centerline{\bigrm and the geometry of minbus} 

\bigskip\bigskip

\centerline{J. Bouttier and E. Guitter}
  \smallskip
  \centerline{Institut de Physique Th\'eorique}
  \centerline{CEA, IPhT, F-91191 Gif-sur-Yvette, France}
  \centerline{CNRS, URA 2306}
\centerline{\tt jeremie.bouttier@cea.fr}
\centerline{\tt emmanuel.guitter@cea.fr}

  \bigskip

     \bigskip\bigskip

     \centerline{\bf Abstract}
     \smallskip
     {\narrower\noindent
We present a detailed calculation of the distance-dependent two-point function 
for quadrangulations with no multiple edges. Various discrete observables 
measuring this two-point function are computed and analyzed in the limit of 
large maps. For large distances and in the scaling regime, we recover the same 
universal scaling function as for general quadrangulations. We then explore the
geometry of ``minimal neck baby universes'' (minbus), which are the outgrowths 
to be removed from a general quadrangulation to transform it into a 
quadrangulation with no multiple edges, the ``mother universe''. We give a 
number of distance-dependent characterizations of minbus, such as the two-point
function inside a minbu or the law for the distance from a random point to the 
mother universe.
\par}

     \bigskip

\nref\QGRA{V. Kazakov, {\it Bilocal regularization of models of random
surfaces}, Phys. Lett. {\bf B150} (1985) 282-284; F. David, {\it Planar
diagrams, two-dimensional lattice gravity and surface models},
Nucl. Phys. {\bf B257} (1985) 45-58; J. Ambj\o rn, B. Durhuus and J. Fr\"ohlich,
{\it Diseases of triangulated random surface models and possible cures},
Nucl. Phys. {\bf B257} (1985) 433-449; V. Kazakov, I. Kostov and A. Migdal
{\it Critical properties of randomly triangulated planar random surfaces},
Phys. Lett. {\bf B157} (1985) 295-300.}
\nref\DGZ{for a review, see: P. Di Francesco, P. Ginsparg and 
J. Zinn--Justin, {\it 2D Gravity and Random Matrices},
Physics Reports {\bf 254} (1995) 1-131.}
\nref\MARMO{J. F. Marckert and A. Mokkadem, {\it Limit of normalized
quadrangulations: the Brownian map}, Annals of Probability {\bf 34(6)}
(2006) 2144-2202, arXiv:math.PR/0403398.}
\nref\LEGALL{J. F. Le Gall, {\it The topological structure of scaling limits 
of large planar maps}, invent. math. {\bf 169} (2007) 621-670,
arXiv:math.PR/0607567.}
\nref\LGP{J. F. Le Gall and F. Paulin,
{\it Scaling limits of bipartite planar maps are homeomorphic to the 2-sphere},
Geomet. Funct. Anal. {\bf 18}, 893-918 (2008), arXiv:math.PR/0612315.}
\nref\MierS{G. Miermont, {\it On the sphericity of scaling limits of 
random planar quadrangulations}, Elect. Comm. Probab. {\bf 13} (2008) 248-257, 
arXiv:0712.3687 [math.PR].}
\nref\LEGALLGEOD{J.-F. Le Gall, {\it Geodesics in large planar maps and 
in the Brownian map}, Acta Math., to appear, arXiv:0804.3012 [math.PR].}
\nref\THREEPOINT{J. Bouttier and E. Guitter, {\it The three-point function 
of planar quadrangulations}, J. Stat. Mech. (2008) P07020, 
arXiv:0805.2355 [math-ph].}
\nref\LOOP{J. Bouttier and E. Guitter, {\it Confluence of geodesic paths and 
separating loops in large planar quadrangulations}, 
J. Stat. Mech. (2009) P03001, arXiv:0811.0509 [math-ph].}
\nref\ADJ{J. Ambj\o rn, B. Durhuus and T. Jonsson, {\it Quantum Geometry:
A statistical field theory approach}, Cambridge University Press, 1997.}
\nref\AW{J. Ambj\o rn and Y. Watabiki, {\it Scaling in quantum gravity},
Nucl.Phys. {\bf B445} (1995) 129-144, arXiv:hep-th/9501049.}
\nref\AJW{J. Ambj\o rn, J. Jurkiewicz and Y. Watabiki,
{\it On the fractal structure of two-dimensional quantum gravity},
Nucl.Phys. {\bf B454} (1995) 313-342, arXiv:hep-lat/9507014.}
\nref\GEOD{J. Bouttier, P. Di Francesco and E. Guitter, {\it Geodesic
distance in planar graphs}, Nucl. Phys. {\bf B663}[FS] (2003) 535-567, 
arXiv:cond-mat/0303272.}
\nref\GW{Z. Gao and N.C. Wormald, {\it The size of the largest components in 
random planar maps}, SIAM J. Discrete Math. {\bf 12} (1999), 217-228.}
\nref\BaFlScSo{C. Banderier, P. Flajolet, G. Schaeffer and M. Soria, {\it 
Random Maps, Coalescing Saddles, Singularity Analysis, and Airy Phenomena},
Random Structures and Algorithms {\bf 19} (2001) 194-246.}
\nref\JainMa{S. Jain and S. Mathur, {\it World-sheet geometry and
baby universes in 2D quantum gravity}, Phys. Lett. {\bf B 286} (1992) 239-246, 
 arXiv:hep-th:9204017.}
\nref\MS{M. Marcus and G. Schaeffer, {\it Une bijection simple pour les
cartes orientables} (2001), available at 
\hhref{http://www.lix.polytechnique.fr/Labo/Gilles.Schaeffer/Biblio/};
see also G. Schaeffer, {\it Conjugaison d'arbres
et cartes combinatoires al\'eatoires}, PhD Thesis, Universit\'e 
Bordeaux I (1998) and G. Chapuy, M. Marcus and G. Schaeffer, 
{\it A bijection for rooted maps on orientable surfaces}, 
SIAM J. Discrete Math. {\bf 23}(3) (2009) 1587-1611, 
arXiv:0712.3649 [math.CO].}
\nref\SCHPRIV{G. Chapuy and G. Schaeffer, private communication.}
\nref\TutteCPM{W.T. Tutte, {\it A Census of Planar Maps}, Canad. J. of Math. 
{\bf 15} (1963) 249-271.}
\nref\Bro{W.G. Brown, {\it Enumeration of nonseparable planar maps},
Canad. J. Math. {\bf 15} (1963) 526-554.}
\nref\BT{W. G. Brown and W.T. Tutte, {\it On the enumeration of rooted
nonseparable planar maps} Canad. J. Math. {\bf 16} (1964) 572-577.}
\nref\ScJa{G. Schaeffer and B. Jacquard, {\it A bijective census of
nonseparable planar maps}, J. Comb. Theory, Ser. A {\bf 83}(1) (1998) 1-20.}
\nref\MOB{J. Bouttier, P. Di Francesco and E. Guitter. {\it 
Planar maps as labeled mobiles},
Elec. Jour. of Combinatorics {\bf 11} (2004) R69, arXiv:math.CO/0405099.}
\newsec{Introduction}

\subsec{The problem}

The study of random maps is an active field of research which raises beautiful 
combinatorial and probabilistic problems. In particular, maps are used in 
physics as discrete models for fluctuating surfaces in a wide range of 
domains, like the study of biological membranes or string theory. Random maps 
can be equipped with statistical models, such as Ising spins, dimers, hard 
particles, and give rise to a large variety of critical phenomena, described 
in the physics literature by the so-called two-dimensional quantum gravity 
[\xref\QGRA,\xref\DGZ]. For large maps, several sensible scaling limits of 
continuous surfaces 
can be reached, depending on the universality class of the model at hand. 
Of particular interest is the so-called Brownian map 
[\xref\MARMO,\xref\LEGALL], which describes the
scaling limit of large non-critical planar maps (in the universality class 
of the so-called pure gravity), like maps with prescribed face degrees, 
for instance planar triangulations (maps with faces of degree 3 only) or
planar quadrangulations (maps with faces of degree 4 only). 

The Brownian map was shown to have the topology of the two-dimensional
sphere [\xref\LGP,\xref\MierS] but it has nevertheless surprising geometrical 
properties, which place 
it half-way between a tree-like object and a smooth surface. For instance,
it was shown that, as would happen in a tree, two geodesic paths leading 
to a given point in the map merge into a common geodesic before 
reaching this point. This is the so-called {\it confluence phenomenon}
\LEGALLGEOD, 
which reveals some underlying tree-like structure of the Brownian map. 
On the other hand, it was shown that the triangle formed by three geodesic paths
linking three points in the map delimit two macroscopic (interior and 
exterior) regions, like in a smooth surface [\xref\THREEPOINT,\xref\LOOP]. 
Heuristically,
it was claimed that large maps can be viewed as made of a large component,
the ``mother universe'' with attached small components, the ``baby universes''
arranged into tree-like structures (see for instance \ADJ). 
A proper definition of mother 
and baby universes is however lacking so far, which prevents from making this 
statement more precise. 

Beside this qualitative picture of the Brownian map, precise quantitative 
measures of its geometry could be obtained, such as the two-point function
[\xref\AW-\xref\GEOD], 
which gives the profile of distances between two random points in the map,
and the three-point function \THREEPOINT, which gives the joint law for the
three distances between three random points. In particular, the two- 
and three-point functions could be computed exactly at the discrete
level in the context of general planar quadrangulations. Here, 
by general quadrangulations, we mean all maps with faces of degree 4
only. These maps cannot have loops (since they are clearly bipartite) but 
they may have {\it multiple edges}, creating cycles of length $2$. 
Each of these cycles acts as a neck separating the quadrangulation 
into two components. Now we may imagine cutting the map along
all its cycles of length $2$, thus disconnecting it into several pieces.
It was shown [\xref\GW,\xref\BaFlScSo] that, 
for quadrangulations with a large number $n$
of faces, exactly one of these pieces has a size ($=$ number of faces) 
of order $n$, while all the others have sizes negligible with respect
to $n$. This provides a proper definition of a mother and baby universes:
the large component constitutes the mother universe. Upon gluing each of the 
cut cycle touching this component into a single edge, the mother universe 
may itself be viewed as a planar quadrangulation which, by construction, 
has no multiple edges. The other pieces, once reglued together, form a 
number of connected branched structures,
each with a boundary of length $2$. Each of these branched structures 
constitutes a baby universe, to be glued by its boundary to the mother universe
to complete the quadrangulation. 

It is well known that general planar (rooted) quadrangulations with $n$ faces 
are in one-to-one correspondence with (rooted) general planar maps with $n$
edges. Under this equivalence, the above decomposition of quadrangulations 
simply corresponds to the well-known decomposition of general maps into
$2$-connected components upon splitting them at their separating vertices. 

It is clear that we have here a very restrictive definition of baby universes, 
the so-called {\it minimal neck baby universes} \JainMa, hereafter 
abbreviated into 
{\it minbus} and corresponding to cutting the map along necks of minimal size 
$2$. Unfortunately, the overlaps between cycles of larger sizes prevent from 
defining a canonical decomposition into more general baby universes. 

Rather than considering general quadrangulations, we may at first
start with the more restricted class of quadrangulations with no multiple edges
(in one-to-one correspondence with $2$-connected maps via the above-mentioned
equivalence). These quadrangulations have no cycles of length $2$ by 
definition, hence are reduced to their mother universe, with no minbus. 
It is expected that, for large distances, the two- and three-point functions of 
these quadrangulations with no multiple edges be essentially the same as 
those of general maps since the presence of minbus should affect only small 
distances.

The purpose of this paper is twofold. In a first part, we present a detailed 
calculation of the {\it two-point function for quadrangulations with no 
multiple edges}. Expressions are given for various observables measuring this 
two-point function at the discrete level and then analyzed in the limit
of large maps. Although different from that of general quadrangulations at 
finite distances, they give the same two-point function at large distances 
and in particular in the universal scaling regime. 
In a second part, we use the connection between general quadrangulations
and quadrangulations with no multiple edges inherited from the above 
decomposition into mother universe and minbus to explore a number of
{\it distance-dependent characterizations of the minbus} themselves in general 
quadrangulations.  We compute for instance the {\it two-point function inside 
a minbu} or the law for the distance from a random point to the mother 
universe. 

The paper is organized as follows. In the second part of 
Section 1, we recall known results on the two-point function 
for general quadrangulations. These results are based on the 
Schaeffer bijection which encodes quadrangulations by so-called well-labeled 
trees. In Section 2.1 we introduce the more restricted family of
{\it well-balanced} trees which code for quadrangulations with no multiple 
edges. Generating functions for these well-balanced trees are then computed 
either directly by solving the appropriate recursion relation (Section 2.2), 
or indirectly via a substitution procedure relating them to generating 
functions for regular well-labeled trees (Section 2.3). We end Section 2 
by deriving expressions for various quantities measuring, at the discrete 
level, the two-point function of quadrangulations with no multiple edges
(Section 2.4). These results are used in Section 3 to address various questions 
on the two-point distance statistics. We first give explicit enumerations 
for small distances and small sizes in Section 3.1, and discuss in Section 3.2 
the case of finite distances 
in quadrangulations of large size $n$ (the local limit). We then explore in
Section 3.3 the 
scaling limit of large distances ($\propto n^{1/4}$) in large quadrangulations 
and recover the universal two-point function of the Brownian map. Section 4
is devoted to the study of minbus in general quadrangulations. We
first recall in Section 4.1 how to cut a general quadrangulation along
minimal necks so as to decompose it into a mother universe and minbus. 
We then explain in Section 4.2 the precise meaning of our generating functions
for well-balanced trees in this context. We finally obtain in Section 4.3 
a number of distance-dependent properties
of minbus such as the probability for two points to lie in the same
minbu as a function of their mutual distance as well as the two-point function
inside a minbu. The law for the distance to the mother universe
or for the number of necks to go through to reach it are derived in Appendix A.
We end this paper by a few concluding remarks in Section 5.

\subsec{General quadrangulations and well-labeled trees: reminders}

In the case of general quadrangulations, a fruitful approach to
questions on the distance statistics relies on the Schaeffer bijection
between quadrangulations and well-labeled trees \MS. It consists in a
one-to-one coding of {\it pointed} (i.e.\ with a marked vertex called
the {\it origin}) quadrangulations with $n$ faces by {\it well-labeled
trees} with $n$ edges, i.e.\ plane trees whose vertices carry integer
labels subject to the two conditions:

\item{$\bullet$ \phantom{i}(i)}{the labels of two vertices adjacent in the 
tree differ by at most $1$.
\item{$\bullet$ (ii)}{the minimum label is $1$.}
\par

\noindent The $n+1$ vertices of the well-labeled tree are in
one-to-one correspondence with the $(n+2)-1$ vertices of the
quadrangulation other than the origin, and the label of a vertex in
the tree is nothing but the distance from the associated vertex to the
origin in the quadrangulation.  Moreover, the $2n$ corners of the tree
are in one-to-one correspondence with the $2n$ edges of the
quadrangulation. By {\it corner}, we mean the angular sector between
two consecutive edges around a vertex of the tree, and if that vertex
has label $\ell$ we also say that the corner has label $\ell$. The
corners with label $\ell$ in the tree are in one-to-one correspondence
with the edges of type $(\ell-1)\to\ell$ in the quadrangulation, i.e.\
the edges connecting a vertex at distance $(\ell-1)$ from the origin
to a vertex at distance $\ell$. Finally, the edges of the tree are in
one-to-one correspondence with the faces of the quadrangulation.

A {\it planted} tree is a (plane) tree with a marked corner, whose
label is called the root label. It is also convenient to introduce
{\it almost well-labeled trees} where the condition (ii) is released
into:

\item{$\bullet$ (ii)'}{the minimum label is larger than or equal to $1$.}
\par
\noindent Attaching a weight $g$ per edge, the generating function
$R_\ell\equiv R_\ell(g)$ of planted almost well-labeled trees with
root label $\ell$ satisfies the recursion relation:
\eqn\recurgen{R_\ell={1\over 1-g\, (R_{\ell-1}+R_\ell+R_{\ell+1})}}
for $\ell\geq 1$, with the initial condition $R_0=0$. This relation simply 
states that a tree with root label $\ell$ is fully characterized by
the sequence of its descending subtrees, which are themselves planted almost 
well-labeled trees with root label $(\ell-1)$, 
$\ell$ or $(\ell+1)$. Equation 
\recurgen\ implies in particular the conservation law:
\eqn\conservgen{C_{\ell+1}=C_\ell\quad {\rm with}\ C_\ell\equiv
R_\ell-g R_{\ell-1}R_\ell R_{\ell+1}}
which may also be derived combinatorially \SCHPRIV. Writing
$C_\ell=C_\infty$ leads after some simple manipulations
to the explicit solution \GEOD:
\eqn\expliRell{\eqalign{R_\ell &= R {(1-x^\ell)(1-x^{\ell+3})
\over (1-x^{\ell+1})(1-x^{\ell+2})}\cr
{\rm where}&\ R={1-\sqrt{1-12g}\over 6g}\ {\rm and}\ x+{1\over x}+1={1\over g R^2}\ .\cr}}
Here, we introduced the quantity $R\equiv\lim\limits_{\ell\to \infty}R_\ell$, 
solution of the quadratic equation $R=1+3 g\, R^2$. 
Of particular interest is the case $\ell=1$ for which conditions (ii) and (ii)'
coincide so that $R_1$ is the generating function of well-labeled trees with 
a marked corner with label $1$. From Schaeffer's bijection, this is also the 
generating function of {\it rooted} (i.e.\ with a marked oriented edge) 
quadrangulations with a weight $g$ per face. Writing $C_1=C_\infty$ gives
\eqn\ronevalue{R_1=R-g\, R^3\ .}
If we now wish to address the question of the two-point function, i.e.\ the
law for the distance between two ``points'' (edges or vertices) picked 
at random in a quadrangulations, this can be done by 
considering several possible distance-dependent generating functions. 
For instance, we may consider pointed quadrangulations with, in
addition to their marked origin vertex, a 
marked edge of type $(\ell-1)\to\ell$ with respect to this origin. Let us
denote by $Q_\ell\equiv Q_\ell(g)$ the corresponding generating function
with a weight $g$ per face. In the tree language, it enumerates 
planted well-labeled trees with root label $\ell$. This leads immediately to 
\eqn\Qell{Q_\ell=R_\ell-R_{\ell-1}}
since the condition (ii) is easily restored from the condition 
(ii)' by eliminating the configurations (with root label $\ell$) having a 
minimum label strictly larger than $1$, counted by $R_{\ell-1}$ by a simple 
shift of labels.

Another measure of the two-point distance statistics is via the 
generating function $F_\ell$ of rooted quadrangulations having, in
addition to their marked oriented root edge, a marked vertex
at distance $\ell$ from the origin of the root edge. In the tree language, 
$F_\ell$ is the generating function of planted well-labeled trees with
a root vertex with label $1$ and with an extra marked vertex with label $\ell$.
The tree then consists of a chain of almost well-labeled trees attached
on both sides of a linear spine linking these two vertices. At the 
endpoint of the spine is attached only one subtree with root label $\ell$.
We shall call these configurations {\it vertex-ended chains} and
their generating function reads
\eqn\defFl{F_\ell= \left\{\sum_{k\geq 0}
\sum_{{\rm paths}\ (1=\ell_0,\ell_1,\cdots, \ell_k=\ell)
\atop \ell_i\geq 1,\ \vert\ell_i-\ell_{i-1}\vert\leq 1, i=1,\cdots k} 
\prod_{i=0}^{k-1}
g \left(R_{\ell_i}\right)^2 \right\}\times R_\ell}
where $k$ is the length of the spine.
\fig{A schematic picture of the re-rooting procedure, showing that there 
is a bijection between, on the one hand, rooted maps with a marked vertex 
at distance $\ell$ from the origin of the root edge, and on the other
hand, pointed maps with a marked edge of type $(\ell-1)\to\ell$ or
$\ell\to(\ell+1)$ with respect to the origin vertex.}{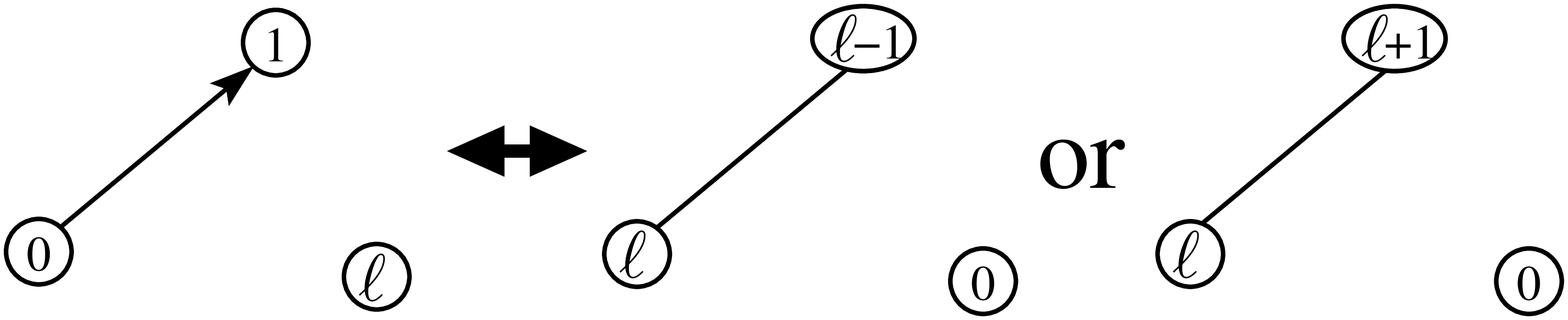}{10.cm}
\figlabel\rerooting
Now, by a simple re-rooting procedure, there is a clear 
bijection between maps with a marked root edge and a marked vertex at 
distance $\ell$ from the origin of the root edge, and maps with a marked 
origin and a marked edge of type $(\ell-1)\to\ell$ or of type $\ell\to(\ell+1)$
with respect to this origin (see figure \rerooting\ for an illustration).
This bijection translates into the simple relation
\eqn\FQrel{F_\ell=Q_\ell+Q_{\ell+1}=R_{\ell+1}-R_{\ell-1}}
for $\ell\geq 1$, a formula which can also be obtained directly by induction
from \defFl\ by use of the conservation law.

Finally, a third interesting measure of the two-point distance statistics 
is via the generating function $H_\ell$ of rooted quadrangulations 
having, in addition to the marked oriented root edge, an extra marked 
edge of type $(\ell-1)\to \ell$ with respect to the origin of the root edge
(if $\ell=1$, we may choose for the extra marked edge the root edge
itself, in which case the configuration is counted with a weight $2$ for 
convenience).
In terms of trees, this amounts to replacing the marked vertex with label
$\ell$ in the configurations counted by $F_\ell$ by a marked corner. 
This leads to the
same chain of almost well-labeled trees as for $F_\ell$, with now two trees 
with label $\ell$ instead of one attached to the endpoint of the spine. 
We shall call these configurations {\it corner-ended chains} and their
generating function reads
\eqn\Hell{H_\ell=R_\ell\ F_\ell=R_{\ell+1}\ R_\ell- R_\ell\ R_{\ell-1}}
for $\ell\geq 1$. 

The generating functions $Q_\ell$, $F_\ell$ and $H_\ell$, although
clearly different, are equivalently good measures of the two-point function 
at the discrete level. In the limit of quadrangulations with a large 
number $n$ of faces and for large distances (of order $n^{1/4}$), they
lead to the same distance statistics, characterized by a unique 
scaling function, the (universal) continuous two-point function. 

\newsec{Quadrangulations with no multiple edges and well-balanced 
well-labeled trees}

We now turn to the study of quadrangulations with no multiple edges,
i.e.\ quadrangulations where {\it all pairs of vertices are linked by at 
most one edge}.  As already mentioned, these (rooted) quadrangulations 
with $n$ faces are in one-to-one correspondence with (rooted) general
maps with $n$ edges {\it having no separating vertex}. Such maps are 
usually called 2-connected or nonseparable and were first enumerated in 
Refs. [\xref\TutteCPM-\xref\BT]. More recently, a bijection between 
nonseparable maps and two different families of trees was discovered 
in Ref.\ScJa, which explains the remarkably simple formula for
their number. These trees are: (1) so-called
{\it description trees} which are a particular class of labeled trees 
(with arbitrary internal degrees) and (2) so-called {\it skew ternary trees} 
which are ternary trees with particular positivity constraints. 

If the enumeration of quadrangulations with no multiple edges is a 
well understood question, much less is known on their distance statistics.
Here we shall present yet another bijection with labeled trees, obtained by
simply restricting the general class of well-labeled trees in 
the Schaeffer bijection to a smaller class of {\it well-balanced} ones
where additional constraints guarantee that the associated quadrangulation 
has no multiple edge. Even if it seems likely that a direct correspondence
can be found between these trees and the description trees of Ref. \ScJa, 
well-balanced trees are particularly adapted to questions involving the 
distance and this is the reason why we use them here.

\subsec{Well-balanced well-labeled trees}
\fig{A schematic picture for the appearance of a double edge when 
reconstructing a quadrangulation from its associated well-labeled tree. 
A double edge is found either (a) when a vertex with label $1$ is not
a leaf, or (b) when the subtree separating two given successive
corners at a vertex with label $\ell>1$ contains no label 
$(\ell-1)$.}{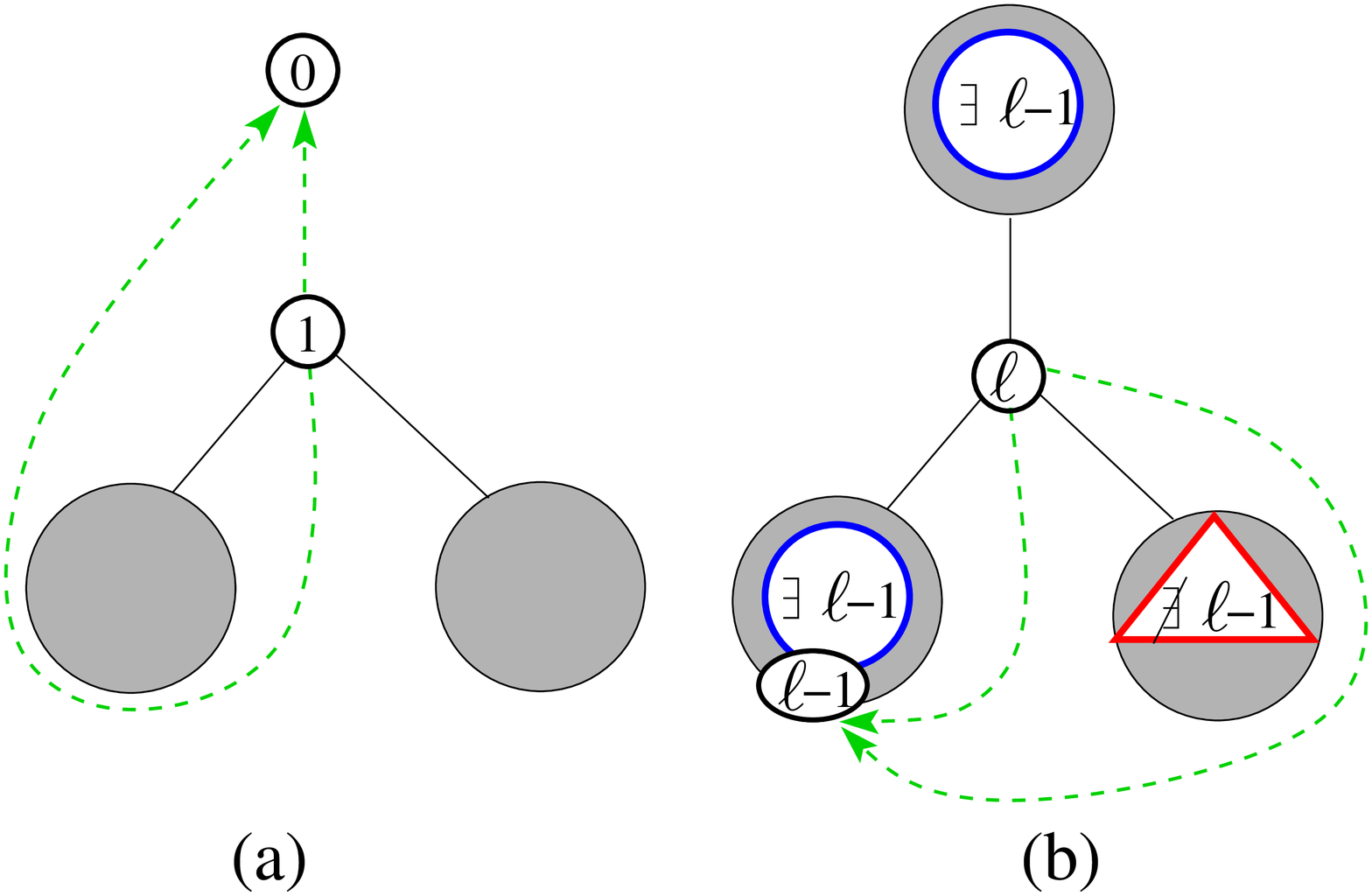}{9.cm}
\figlabel\condition

Before we address the question of multiple edges, let us recall
how we obtain a pointed quadrangulation from its associated 
well-labeled tree in the Schaeffer bijection. First we take 
for the vertices of the quadrangulation all the vertices 
of the tree plus an extra added vertex with label $0$, which will be the 
origin of the quadrangulation.
The edges of the quadrangulation are then obtained by connecting each 
{\it corner} of the tree with label $\ell$ to its {\it successor}, 
which is the first corner at a vertex with label $\ell-1$ encountered clockwise
around the tree if $\ell>1$, or the added vertex with label $0$ if $\ell=1$.
These connections can be performed without edge crossings.
We finally erase all the original tree edges as well as the vertex labels.

It is now straightforward to deduce at which condition the quadrangulation
will have no multiple edges. First, as any corner at a vertex with label
$1$ will be linked to the origin, the absence of multiple edges requires
that there is exactly one corner at any vertex with label $1$
(see figure \condition-(a) for a counterexample), namely that:

\item{$\bullet$ (a)} vertices with label $1$ are leaves of the tree.
\par
\noindent Consider now a vertex with label $\ell>1$, which we assume is not 
a leaf of the tree,
and consider two successive corners clockwise around this vertex.
A multiple edge will appear in the quadrangulation if these two corners have 
the same successor. This occurs if the subtree attached to the considered 
vertex and lying between the two considered corners does not contain any
vertex with label $\ell-1$ (see figure \condition-(b)).
A necessary and sufficient condition for the absence of multiple edge 
of type $\ell\to(\ell-1)$ is that:

\item{$\bullet$ (b)}{each of the $k$ subtrees attached to a vertex of degree
$k$ in the tree and with label $\ell>1$ contains a vertex with label $\ell-1$.}
\par
\noindent Note that this condition is automatically satisfied if the
vertex is a leaf ($k=1$) with label $\ell>1$ as the only attached subtree 
is then the entire tree itself (minus the considered vertex) which, 
from (i) and (ii), contains all integer labels between $1$ and $\ell$.

A well-labeled tree satisfying (a) and (b) will be called 
{\it well-balanced}. We have a bijection between pointed quadrangulations 
with $n$ faces and with no multiple edges and well-balanced well-labeled trees 
with $n$ edges.

\subsec{Generating functions for almost well-balanced trees and 
enumeration of quadrangulations with no multiple edges}
\fig{A schematic picture of the relation $p_1=z(1+r_2)$ for the
generating function $p_1$ of well-balanced well-labeled trees planted
at a vertex with label $1$. This vertex is necessarily a leaf, linked 
either to another leaf with label $1$ or to an almost 
well-balanced well-labeled tree with root label $2$, represented
here by a grey triangle.}{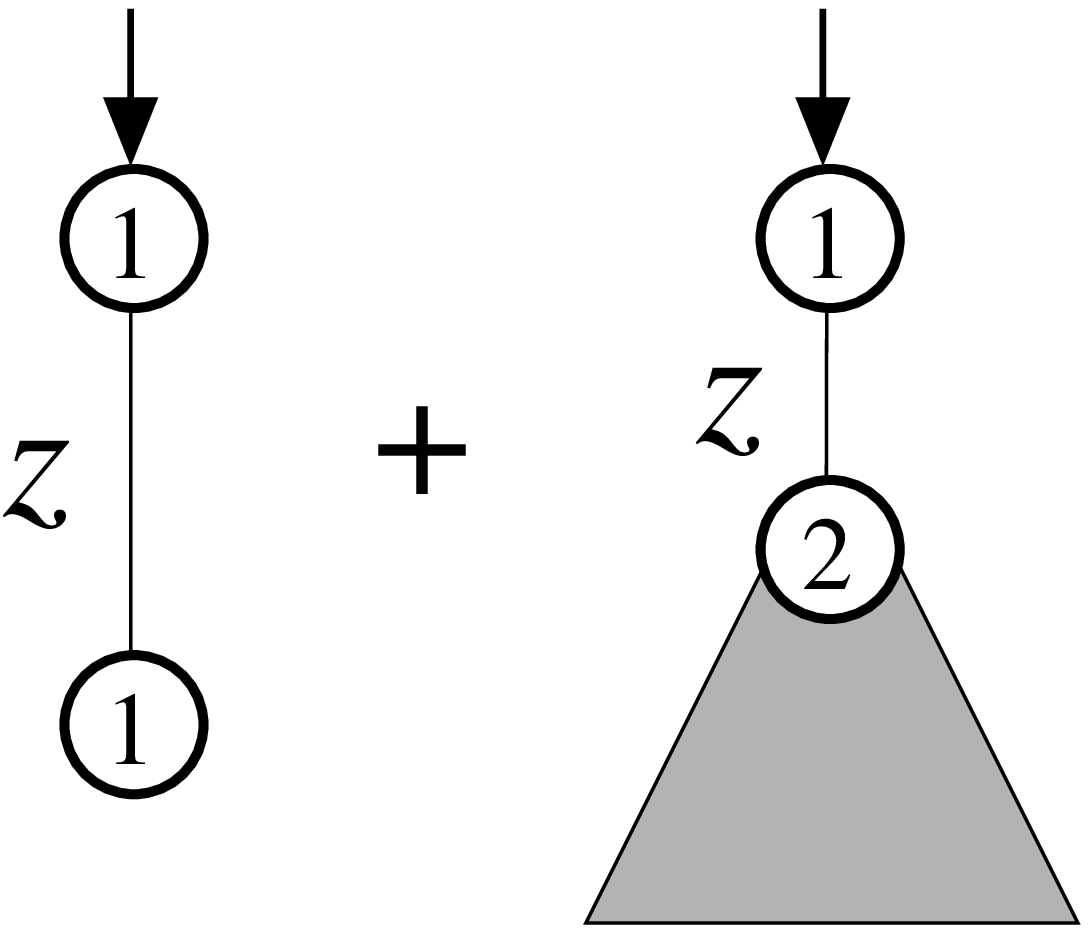}{4.cm}
\figlabel\genforqwnme
As in Section 1.2, let us slightly modify the label constraints on the trees
so as to get simpler generating functions. Let us define {\it almost
well-balanced well-labeled trees} as planted trees which are almost 
well-labeled, i.e.\ carry integer vertex labels satisfying (i) and 
(ii)' above, and are also almost well-balanced, i.e.\ satisfy the 
conditions:
\item{$\bullet$ (a)'}{vertices with label $1$ have no {\it descending}
subtrees.}
\item{$\bullet$ (b)'}{any {\it descending} subtree attached to a vertex with 
label $\ell'> 1$ in the tree contains a vertex with label $(\ell'-1)$.}
\par
\noindent We denote by $r_\ell\equiv r_\ell(z)$ the generating function
of these trees with root label $\ell$ and with a weight $z$ per edge.
If $\ell>1$, the condition (a)' is equivalent to (a) but (b)'
is weaker than (b) as we do not impose any constraint on the ascending subtrees.
Therefore almost well-balanced trees are not well-balanced in general. 
When $\ell=1$, i.e.\ when the root label itself is $1$, (a)' implies 
drastically that the whole tree reduces to a single vertex with 
label $1$, with the trivial generating function 
\eqn\roneval{r_1=1\ .}
This is to be contrasted with the generating function $p_1\equiv p_1(z)$ 
of the truly well-balanced well-labeled trees planted at a vertex 
with label $1$. This generating function, which is also that of {\it rooted} 
quadrangulations with no multiple edges with a weight $z$ per face, 
can be obtained as follows. 
From (a), the root 
vertex in this case is necessarily a leaf and either the tree reduces to two 
leaves with label $1$ connected by a single edge (see figure \genforqwnme-(a)) 
or the tree is made of a leaf with label $1$ connected by an edge to an 
almost well-labeled tree planted at a vertex with label $2$ (see figure
\genforqwnme-(b)). It is enough to demand that this attached tree be 
almost well-balanced as, if so, the entire tree is well-balanced. This
is because condition (b) is also clearly satisfied for ascending subtrees 
in this case, due to the existence of a label $1$ at the top of the tree, 
which, from (i), ensures that any vertex with label $\ell>1$ has an 
ancestor vertex in the tree with label $(\ell-1$). To summarize, 
the generating function of rooted quadrangulations with no multiple edges
reads:
\eqn\valq{p_1=z+z\, r_2\ .}

\fig{Top: A schematic picture of the relation (2.3) for the
generating function $r_\ell$ of almost well-balanced well-labeled trees
with root label $\ell$. Any such tree is entirely specified by
the sequence of its descending subtrees (grey triangles), which are
themselves almost well-balanced well-labeled trees with root label 
$(\ell-1)$, $\ell$ or $(\ell+1)$. Each of these subtrees must contain 
a vertex with label $(\ell-1)$. Bottom: A schematic representation of
the relation (2.4).}{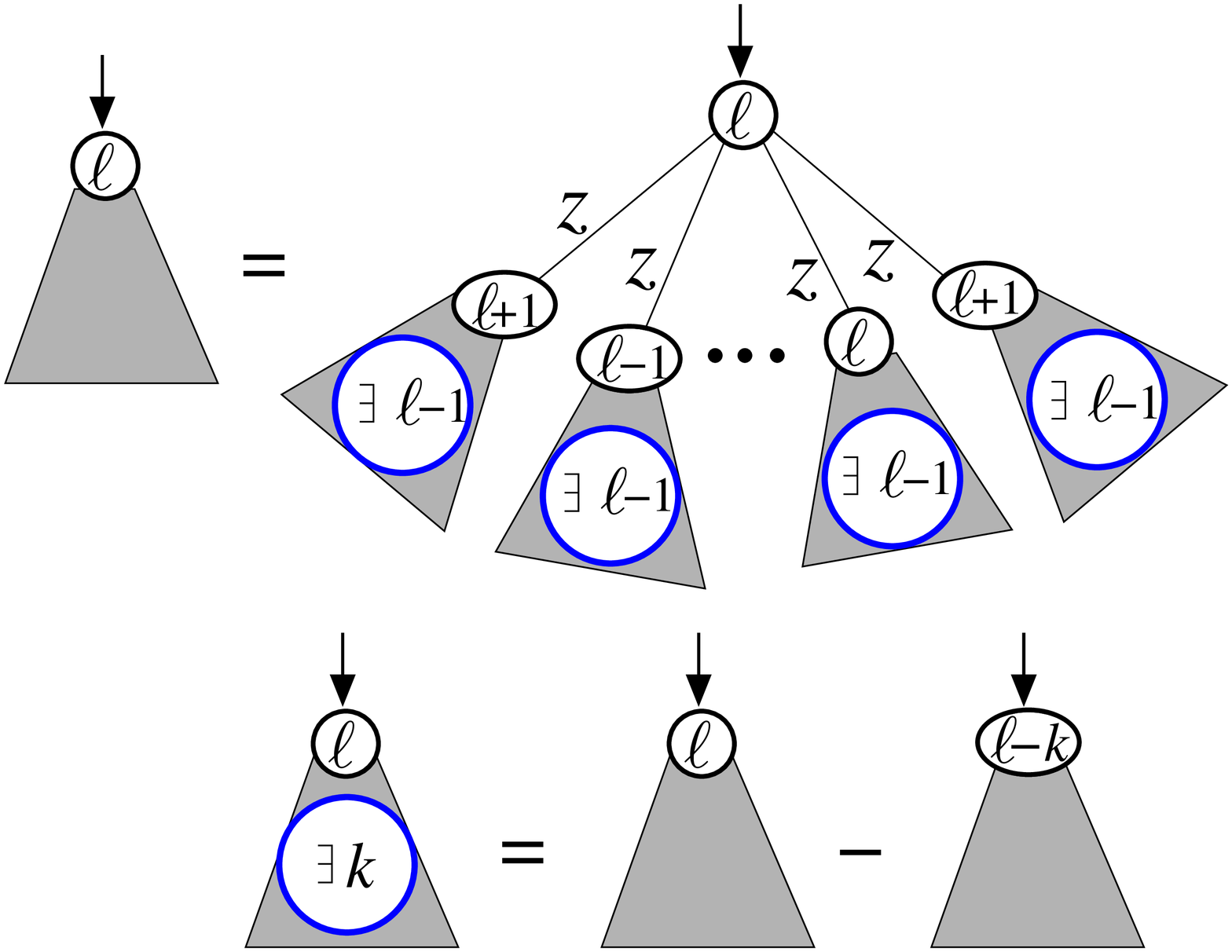}{8.cm}
\figlabel\recursion
Let us now derive an explicit expression for the generating functions $r_\ell$.
It may be obtained in two different ways: we may either get it as the solution
of some recursion relation of the type \recurgen, now incorporating the 
constraint of having almost well-balanced trees. This approach is presented 
in details in this Section. Or, as explained in the next Section, we may 
obtain it directly from the explicit form \expliRell\ of $R_\ell$ via a simple 
substitution procedure.  

The generating functions $r_\ell$ obey the following relation
for $\ell\geq 1$, illustrated in figure \recursion:
\eqn\recurone{r_\ell= {1\over 1-z (r_{\ell-1}^{(\ell-1)}+r_\ell^{(\ell-1)}
+r_{\ell+1}^{(\ell-1)})}}
where we introduced the generating 
function $r_\ell^{(k)}$ ($0\leq k \leq \ell$) for almost 
well-balanced well-labeled trees planted at a vertex with 
label $\ell$ and satisfying the extra requirement that they contain a vertex
with label $k$. Equation \recurone\ simply states that an almost well-balanced
well-labeled tree with root label $\ell$ may be viewed as a sequence of almost 
well-balanced well-labeled subtrees with root labels $(\ell-1)$, $\ell$
or $(\ell+1)$, each of them containing a vertex with label $(\ell-1)$.

Now any almost well-balanced well-labeled tree with root label
$\ell$ and with no occurrence of the label $k\leq \ell$ has all its labels
strictly larger than $k$. Shifting all labels by $k$ creates a tree
which is still almost well-balanced and almost well-labeled, and now
has root label $\ell-k$. This leads to the relation
\eqn\rlk{r_\ell^{(k)}=r_\ell-r_{\ell-k}}
with $r_0=0$. We have in particular $r_{\ell-1}^{(\ell-1)}=r_{\ell-1}$
since a label $(\ell-1)$ is already present at the level of the root,
$r_{\ell}^{(\ell-1)}=r_{\ell}-r_1=r_\ell-1$ since a label $(\ell-1)$ is
automatically present in each subtree and therefore the only situation
with no label $(\ell-1)$ is when the tree reduces to its root vertex.
Finally, we have $r_{\ell+1}^{(\ell-1)}=r_{\ell+1}-r_2$ so that 
equation \recurone\ translates into the recursion relation 
\eqn\recur{r_\ell= {1\over 1-z (r_{\ell-1}+r_\ell+r_{\ell+1})+z (r_1+r_2)}}
valid for all $\ell\geq 1$. Taking $\ell\to \infty$ gives
\eqn\recurinfty{r\equiv \lim_{\ell\to \infty}r_\ell=
{1\over 1-3\, z\, r + z + z\, r_2}}
namely
\eqn\rtwo{r_2=3r -1 +{1-r\over z\, r}\ .}
Writing \recur\ for $\ell$ and $\ell+1$ as
\eqn\rerecur{\eqalign{
1 & = r_\ell(1+z (r_1+r_2))-z r_\ell(r_{\ell-1}+r_\ell+r_{\ell+1})\cr
1 & = r_{\ell+1}(1+z (r_1+r_2))-z r_{\ell+1}(r_\ell+r_{\ell+1}+r_{\ell+2})
\ ,\cr}}
multiplying the first line by $r_{\ell+1}$, the second line by $r_\ell$
and taking the difference leads to the conservation property
\eqn\conservone{c_{\ell+1}=c_\ell \ \ {\rm with}\ c_\ell\equiv
r_\ell-z r_{\ell-1} r_\ell r_{\ell+1}\ .}
Alternatively, this conservation property may be obtained by the same
combinatorial argument as that for general well-labeled trees
\SCHPRIV: let us consider planted almost well-balanced well-labeled
trees having root label $\ell$ and {\it containing a vertex with label
$1$}. By definition, their generating function is
$r_{\ell}^{(1)}$. But also any such tree can be
decomposed into three parts as follows. In the sequence of subtrees
attached to the root, let us single out the leftmost one containing a
label $1$, which is the first part in our decomposition. The second
(resp.\ third) part is formed by all subtrees on its left (resp.\
right) together with the root vertex. It is easily seen that the three
parts are planted almost well-balanced well-labeled trees. Moreover,
the first part has root label $\ell-1$, $\ell$ or $\ell+1$ and
necessarily contains a label $1$, the second part contains no label
$1$ therefore by shifting all labels by $1$ we obtain a tree with root
label $\ell -1$, finally the third tree has root label $\ell$ and
obeys no further constraint. Clearly this decomposition is reversible
and the total number of edges is decreased by one, leading to the
relation:
\eqn\combiconserv{r_{\ell}^{(1)} = z \left( r_{\ell-1}^{(1)} + r_{\ell}^{(1)}
+ r_{\ell+1}^{(1)} \right) r_{\ell-1} r_{\ell}}
which, together with \rlk, immediately
implies the conservation property \conservone.

We now deduce that $c_\ell=c_1=1$ for all $\ell\geq 1$, namely
\eqn\conserv{r_\ell=1+z r_{\ell-1}r_\ell r_{\ell+1}}
and in particular
\eqn\eqnforr{r=1+z r^3}
or explicitly
\eqn\valr{r={2\over \sqrt{3z}} \sin\left({1\over 3} \arcsin 
\sqrt{27z\over 4}\right)=\sum_{n\geq 0}{(3n)!\over (2n+1)! n!} z^n\ .}
From \rtwo, we deduce that 
\eqn\valrtwo{r_2=3r-1-r^2=1+\sum_{n\geq 1} {2\, (3n)!\over (2n+1)!(n+1)!} 
z^n\ .}
From \valq, the generating function $p_1$ reads finally
\eqn\valqbis{p_1=z\, r\, (3-r)=\sum_{n\geq 1} {2\, (3n-3)!\over (2n-1)!n!} z^n}
from which we read off the number of rooted quadrangulations with $n$ faces
and with no multiple edges [\xref\TutteCPM-\xref\BT]. Here we observe
a connection with the formulation of Ref.\ScJa\ by noting that $r$ 
(respectively $p_1$) is precisely the generating function
for ternary (respectively skew ternary) trees. 

The conservation law \conservone\ is identical to that, \conservgen, of
general well-labeled trees and only the precise value of the conserved 
quantity differs ($1$ instead of $R_1$). Repeating the manipulations 
leading from \conservgen\ to the explicit form \expliRell, 
we immediately deduce from the new conservation law \conservone\ the form of 
$r_\ell$ for general $\ell$:
\eqn\valrl{r_\ell=r\, {(1-y^\ell)(1-y^{\ell+3})\over (1-y^{\ell+1})
(1-y^{\ell+2})}}
with $r$ as above and $y\equiv y(z)$ solution of
\eqn\eqnforx{y+{1\over y}+1={1\over z r^2}\ .}

\subsec{Approach by substitution}
Rather than solving the recursion relation \recur, we may
alternatively recover the above expression for $r_\ell$, as well as 
all the results of Section 2.2, directly from the known form \expliRell\
of $R_\ell$ via a substitution procedure as follows.

Starting with an almost well-labeled tree planted at a vertex with label
$\ell$, as counted by $R_\ell$, we may realize conditions (a)'-(b)' 
by simply erasing all descending subtrees attached to vertices with label $1$
and, at each vertex with label $\ell'>1$, erase any of its descending 
subtrees which does not contain a label $\ell'-1$. The remaining tree
is clearly an almost well-balanced well-labeled tree, counted by $r_\ell$. 
Conversely, we recover a general almost well-labeled tree from 
an almost well-balanced well-labeled one by attaching {\it at each corner}
with label $\ell'\geq 1$ an arbitrary well-labeled subtrees with root label 
$\ell'$ and with no label $\ell'-1$, i.e. with all its labels larger than 
or equal to $\ell'$. Any such subtree is counted by $R_1$, independently of 
$\ell'$, as obtained by shifting all labels by $\ell'-1$. For a planted tree 
with $n$ edges, there are exactly $2n+1$ corners where to attach the subtrees.
This results into the identity
\eqn\rtoR{R_\ell(g)=R_1(g)\, r_\ell\left(g \left(R_1\left(g\right)\right)^2 
\right)\ .}
Introducing the function $z(g)$ defined as
\eqn\zgz{z(g)\equiv g\ (R_1(g))^2 }
and its inverse function $g(z)$, we may write \rtoR\ as
\eqn\rtoRbis{
r_\ell(z)= {R_\ell(g(z))\over R_1(g(z))}\quad \hbox{and in particular}\quad 
r(z)= {R(g(z))\over R_1(g(z))}\ .
}
Using $R_1=R-g R^3$, we may write $R/R_1=1+g R_1^2 (R/R_1)^3$ for $g=g(z)$, 
leading to the characterization $r=1+ z r^3$ of $r\equiv r(z)$, which
matches the characterization \eqnforr\ of previous section. 
Combining \zgz\ and \rtoRbis, we deduce that
\eqn\zrtwo{z\, r(z)^2 = g(z)\, R((g(z))^2}
so that, if $x=x(g)$ is the solution of $x+1/x+1=1/(g R^2)$, 
then $y\equiv y(z)= x(g(z))$ is the 
solution of $y+1/y+1=1/(z r^2)$, while \rtoRbis\ gives explicitly 
\eqn\rter{r_\ell(z)={R(g(z))\over R_1(g(z))}\ {
\left(1-\left(x(g(z)\right)^\ell \right)
\left(1-\left(x(g(z)\right)^{\ell+3} \right)\over
\left(1-\left(x(g(z)\right)^{\ell+1} \right)
\left(1-\left(x(g(z)\right)^{\ell+2} \right)} = r\, 
{(1-y^\ell)(1-y^{\ell+3})\over (1-y^{\ell+1})(1-y^{\ell+2})}}
which is precisely the expression \valrl.

\subsec{Two-point function for quadrangulation with no multiple edges}

In the case of general quadrangulations, we could easily go from almost
well-labeled trees (as counted by $R_\ell$) to fully well-labeled ones (as
counted by $Q_\ell$, $F_\ell$ or $H_\ell$). Using \Qell, \FQrel\ or \Hell,
we could extract from \expliRell\ explicit expressions for the various
discrete versions of the two-point function. For quadrangulations with no 
multiple edges, 
we may again restore the condition (ii) of well-labeled trees from the 
condition (ii)' of almost well-labeled ones by considering the analog
of $Q_\ell$, i.e.\ the generating
function $q_\ell\equiv q_\ell(z)$ defined as
\eqn\qell{q_\ell= r_\ell-r_{\ell-1}\ .}
Unfortunately, the trees counted by $q_\ell$ are in general not well-balanced 
and restoring 
the condition (b) from the condition (b)' is not so simple as it 
requires considering ascending subtrees. As such, the knowledge of 
$r_\ell$ or $q_\ell$ is not directly sufficient to answer the question of the 
two-point function. 

Fortunately, we may circumvent this problem by considering instead the analogs 
of the generating functions $F_\ell$ and $H_\ell$ 
of Section 1.2.
\fig{A schematic picture of the generating functions $f_\ell$, 
$h_\ell$, $e_\ell$ and $g_\ell$ for chains of almost well-balanced
well-labeled trees. The end of the chain is either a marked  
vertex with label $\ell$ (in $f_\ell$ or $e_\ell$) or a marked corner with
label $\ell$ (in $h_\ell$ or $g_\ell$).
 We ensure that the trees enumerated 
by $g_\ell$ or $e_\ell$ are fully well-balanced by forbidding the appearance 
of labels $1$ along the spine of the chain and by demanding that, for each 
vertex with label $k$ along the spine, the subtree formed by the part of 
the spine lying strictly below this vertex and by the attached subtrees 
contains a label $(k-1)$.
}{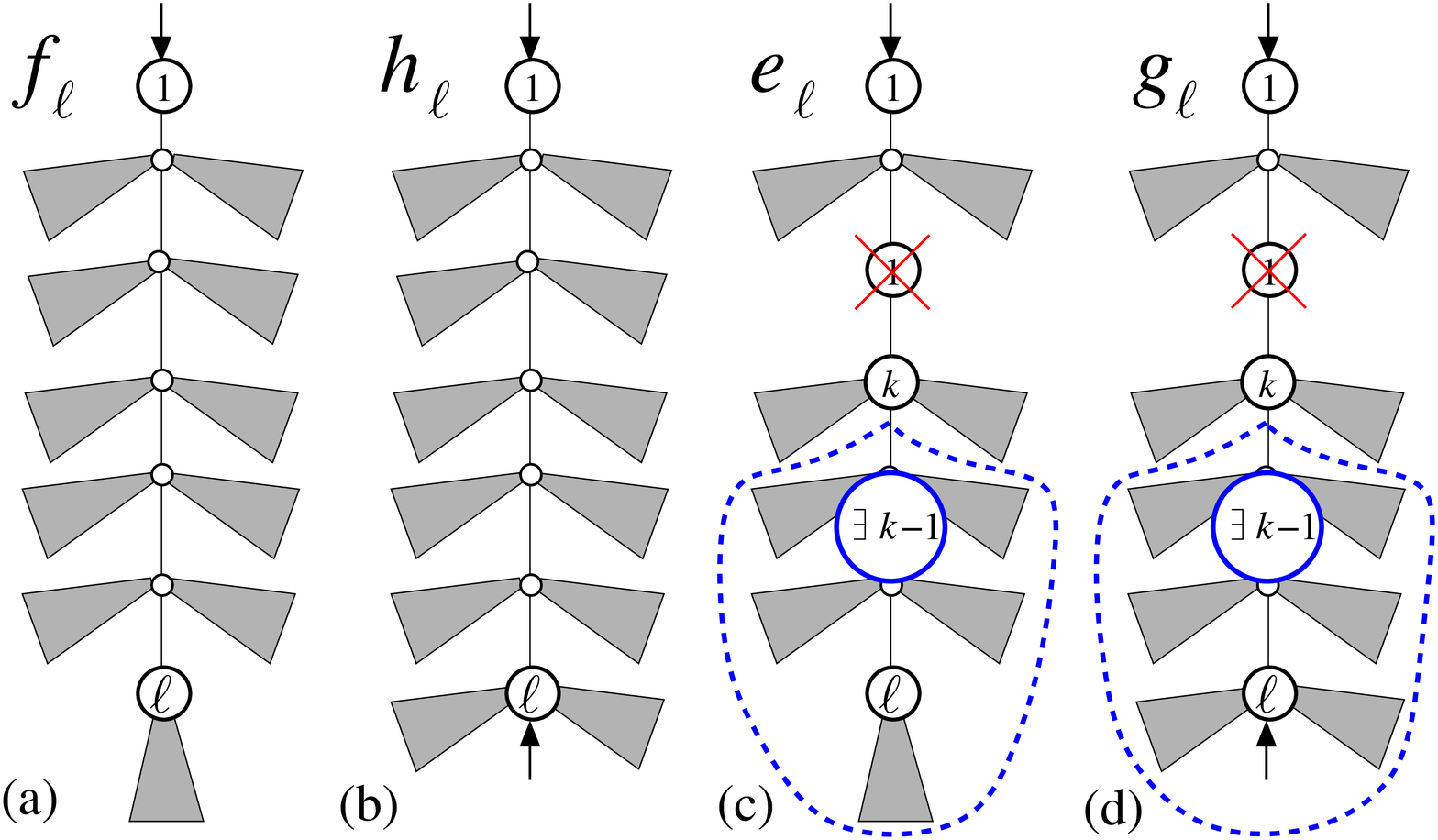}{11.cm}
\figlabel\efgh 
More precisely, as in \defFl, we define the generating function 
$f_\ell\equiv f_\ell(z)$ for {\it almost well-balanced vertex-ended chains} as 
\eqn\defflhl{f_\ell= \left\{\sum_{k\geq 0}
\sum_{{\rm paths}\ (1=\ell_0,\ell_1,\cdots, \ell_k=\ell)
\atop \ell_i\geq 1,\ \vert\ell_i-\ell_{i-1}\vert\leq 1, i=1,\cdots k} 
\prod_{i=0}^{k-1}
z \left(r_{\ell_i}\right)^2 \right\}\times r_\ell}
and the generating function $h_\ell\equiv h_\ell(z)$ for
{\it almost well-balanced corner-ended chains}:
\eqn\fhlink{h_\ell= r_\ell\, f_\ell}
with one extra subtree with root label $\ell$ (see figure \efgh-(a) and (b)
for an illustration).
The generating function $f_\ell$ satisfies the recursion
\eqn\recurfl{f_\ell=\delta_{\ell,1}+ 
z\, r_\ell\, (f_{\ell-1}r_{\ell -1}+f_\ell r_\ell+
f_{\ell+1}r_{\ell+1})}
with $f_0=0$, which determines all $f_\ell$ with $\ell\geq 1$ 
as power series in $z$ from the initial condition 
$f_\ell=z^{\ell-1}+{\cal O}(z^\ell)$.
Writing the conservation law \conservone\ as 
$c_{\ell+1}=c_{\ell-1}+\delta_{\ell,1}$ (valid is this form also when 
$\ell=1$), we have
\eqn\guessf{\eqalign{r_{\ell+1}-r_{\ell-1}& = \delta_{\ell,1}+ 
z r_\ell (r_{\ell+1} r_{\ell+2}-
r_{\ell-1}r_{\ell-2})\cr & = \delta_{\ell,1}+
 z r_\ell (r_{\ell+1}(r_{\ell+2}-r_{\ell})+r_\ell(r_{\ell+1}-r_{\ell-1})+
r_{\ell-1}(r_\ell-r_{\ell-2}))\cr}}
from which we identify the solution of \recurfl\ as
\eqn\valfl{f_\ell=r_{\ell+1}-r_{\ell-1}
=r\, y^{\ell-1} {(1-y)(1-y^2)^2
(1-y^{2\ell+3})\over (1-y^\ell)(1-y^{\ell+1})(1-y^{\ell+2})(1-y^{\ell+3})}} 
while
\eqn\valhl{h_\ell=r_{\ell+1}r_\ell-r_\ell r_{\ell-1}
=r^2\, y^{\ell-1} {(1-y)(1-y^2)^2
(1-y^{2\ell+3})\over (1-y^{\ell+1})^2(1-y^{\ell+2})^2}\ .} 
We therefore have the same relations \valfl\ and \valhl\ for the generating
function of our chains of well-balanced trees in terms of $r_\ell$ as
those \FQrel\ and \Hell\ we had for general trees. 
This is not a surprise as, by a 
simple substitution argument as above, we may write directly
\eqn\substFH{
F_\ell(g)=R_1(g)\ f_\ell\left(g \left(R_1\left(g\right)\right)^2 \right)\ ,\quad
H_\ell(g)=\left(R_1\left(g\right)\right)^2
\ h_\ell\left(g \left(R_1\left(g\right)\right)^2 \right)\ .
}
so that \FQrel\ and \Hell\ immediately translate into \valfl\ and \valhl\
upon taking $g=g(z)$.

The almost well-balanced chains above are not fully well-balanced trees
in general but this problem may now be cured by a simple inclusion-exclusion
procedure. More precisely, in the trees counted by $f_\ell$ or $h_\ell$, 
the condition (b) 
is automatically satisfied for all ascending subtrees, due to the presence of 
the label $1$ at
the top of the tree which, from (i), guarantees that any vertex $\ell'>1$ 
in the tree has an ancestor with label $\ell'-1$. Since we attached almost 
well-balanced subtrees to vertices of the spine, condition (b) is also
satisfied for all descending subtrees {\it except possibly at vertices of
the spine itself} and only for the descending subtree consisting of the part 
of the spine lying strictly below such a spine vertex, together with the 
attached subtrees (see figure \efgh\ for an illustration). 
Similarly, the condition (a) is satisfied everywhere except possibly 
along the spine where we may encounter bivalent vertices with label $1$.

Let us denote by $e_\ell\equiv e_\ell(z)$ 
(respectively $g_\ell\equiv g_\ell(z)$) the generating function 
for fully {\it well-labeled vertex-ended chains} (respectively 
{\it well-labeled corner-ended chains}) with the conditions (a) and (b) 
also satisfied 
along the spine (see figure \efgh-(c) and (d) for an illustration). We
explain below how to obtain $e_{\ell}$ and $g_\ell$ from $f_\ell$ and
$h_\ell$ but let us first observe that,
from the bijection of Section 2.1, $e_\ell$ and $g_\ell$
are good generating functions for quadrangulations with no multiple edges. 
More precisely, $e_\ell$ is the generating
function of {\it rooted} quadrangulations with no multiple edges with
a weight $z$ per face and, in addition to their marked 
oriented root edge, with a marked vertex at distance 
$\ell$ from the origin of the root edge. 
Similarly, $g_\ell$ is precisely the generating function of 
{\it rooted} quadrangulations with no multiple edges and in addition with
their marked oriented root edge, with an extra marked edge of type 
$(\ell-1)\to \ell$,
i.e.\ connecting a vertex at distance $\ell-1$ from the origin of the 
root edge to a vertex at distance $\ell$. 

These interpretations hold only for $\ell>1$ while the case $\ell=1$ 
requires more attention: for convenience, if $\ell=1$, we decide 
{\it not to include} in $e_1=g_1$ the contribution $1$ corresponding
to the trivial case where the spine 
(and consequently the entire tree) reduces to a single vertex
with label $1$. In other words, the spine is required to have a non-zero
length. Returning to maps, this implies that $e_1$ counts
rooted quadrangulations with no multiple edges having an extra marked
vertex adjacent to the origin but {\it different from the endpoint of 
the root edge}. 
We may reinstate the missing configurations by considering
the generating function $p_1+e_1$ instead. Similarly, $g_1$ counts rooted 
quadrangulations with no multiple edges having an extra marked edge incident 
to the origin but {\it different from the root edge itself}. Again
we may reinstate the missing configurations by considering $p_1+g_1$. 

Let us now relate $e_\ell$ and $g_\ell$ to $f_\ell$ and $h_\ell$.
By looking, in a configuration counted by $h_\ell$, at the first 
``unsatisfied'' vertex along the spine, with label $k$, we deduce
the relation, valid for $\ell>1$
\eqn\ghlink{h_\ell=g_\ell+\left(\sum_{k=1}^{\ell-1} g_k\, h_{\ell+1-k}\right) 
+ g_\ell (h_1-1) \ \ \ (\ell > 1) \ .}
The first term corresponds to the case where there is no unsatisfied 
vertex, leading to the desired $g_\ell$. The second term corresponds
to a first unsatisfied vertex along the spine with label $k$ between 
$1$ and $\ell-1$: the part of the tree lying above
the unsatisfied vertex (including the left and right subtrees attached to 
this vertex) has no unsatisfied vertex and final label $k$, hence is
counted by $g_k$ while the rest of the tree has no label $k-1$ and
therefore has all its labels strictly larger than $k-1$. Upon shifting
all labels by $k-1$, we get a tree whose root is a leaf with label $1$ 
and whose final vertex at the end of the spine has label $\ell+1-k$, hence
a tree counted by $h_{\ell+1-k}$. This situation incorporates the
case (when $k=1$) of a label $1$ along the spine with an undesired descending
subtree, counted by $h_\ell$. 
The third term corresponds to having the first unsatisfied vertex with
label $\ell$, in which case we can repeat the above argument provided
we make sure that the descending subtree is not empty, resulting in
a factor $h_1-1$ instead of $h_1$. 
Finally, any vertex with label $k>\ell$ along the spine
cannot be unsatisfied due to the presence of the vertex with label $\ell$
at the end of the spine, which implies that all intermediate values of
labels appear inbetween along the spine.
For $\ell=1$, we have the relation
\eqn\ghlinkone{h_1=1+g_1 h_1}
obtained by cutting a configuration counted by $h_1$ at the first 
label $1$ encountered along the spine (and different 
from the root vertex). This relation, together with \ghlink, 
can be summarized into 
\eqn\ghlinkbis{h_\ell=\delta_{\ell,1}+\sum_{k=1}^{\ell} g_k\, h_{\ell+1-k}}
which allows in principle to compute $g_\ell$ from the explicit form
\valhl\ of $h_\ell$.  Introducing a new parameter $t$ conjugate to $\ell$
and the corresponding generating functions
\eqn\hatgenf{
{\hat h}(t,z)\equiv\sum_{\ell\geq 0}h_{\ell+1}(z)\, t^\ell\ ,\quad
{\hat g}(t,z)\equiv\sum_{\ell\geq 0}g_{\ell+1}(z)\, t^\ell\ ,}
equation \ghlinkbis\ may be equivalently written as
\eqn\hatrel{{\hat h}(t,z)=1+{\hat g}(t,z){\hat h}(t,z)\quad {\rm i.e.}\ \ 
{\hat g}(t,z) = 1-{1\over {\hat h}(t,z)}\ .}
As for $e_\ell$, we have the relation, valid for $\ell>1$
\eqn\eflink{f_\ell=e_\ell +\left(\sum_{k=1}^{\ell-1} g_k\, f_{\ell+1-k}\right)
+g_\ell (f_1-1)\ \ \ \ (\ell > 1)}
obtained again by looking at the first unsatisfied vertex along
the spine in a configuration counted by $f_\ell$.
For $\ell=1$, we have instead $f_1=1+g_1 f_1$ while $e_1=g_1$.
We may therefore write 
$f_1=1+(e_1-g_1)+g_1 f_1$ which can be summarized with \eflink\ into
\eqn\eflinkbis{f_\ell=\delta_{\ell,1}+(e_\ell-g_\ell)+
\sum_{k=1}^{\ell} g_k\, f_{\ell+1-k}\ .}
Introducing again
\eqn\hatgenfbis{
{\hat f}(t,z)\equiv\sum_{\ell\geq 0}f_{\ell+1}(z)\, t^\ell\ ,\quad 
{\hat e}(t,z)\equiv\sum_{\ell\geq 0}e_{\ell+1}(z)\, t^\ell}
we may rewrite \eflinkbis\ as
\eqn\hatrelbis{{\hat f}(t,z)=1+({\hat e}(t,z)-{\hat g}(t,z))+
{\hat f}(t,z){\hat g}(t,z)
\quad {\rm i.e.}\ \ 
{\hat e}(t,z) = {{\hat f}(t,z)-1\over {\hat h}(t,z)}}
where we used \hatrel\ to eliminate ${\hat g}(t,z)$.
This again allows us in principle to compute $e_\ell$ from the explicit
forms \valfl\ and \valhl\ of $f_\ell$ and $h_\ell$. 

Let us end this section by considering the generating function 
$p_\ell\equiv p_\ell(z)$ of {\it pointed} quadrangulations 
with no multiple edges, with a weight $z$ per face and with, in
addition to their marked origin vertex, a marked edge of type 
$(\ell-1)\to \ell$ with respect to this origin. As we already discussed, 
we may exchange the role of the marked vertices and edges
(see figure \rerooting\ for an illustration), so that we may write
\eqn\perel{e_\ell=p_\ell+p_{\ell+1}\qquad (\ell>1)}
while, for $\ell=1$, we have $e_1=p_2$ since we imposed in the configurations
counted by $e_1$ that the marked vertex be different from the 
endpoint of the root edge, hence the marked vertex is at distance 
$2$ from this endpoint. Recall finally that $p_1$ counts
rooted quadrangulations. Introducing
\eqn\hatp{{\hat p}(t,z)\equiv \sum_{\ell\geq 0} p_{\ell+1}(z)\, t^\ell}
we may summarize the above relations into
\eqn\eplink{\eqalign{{\hat e}(t,z)& =e_1(z)+
{\hat p}(t,z)-p_1(z)+{{\hat p}(t,z)-t p_2(z) -p_1(z)\over t}
\cr & {\rm i.e.}\ \ {\hat p}(t,z)=p_1(z)+{t\over 1+t} {\hat e}(t,z)= p_1(z)
+{t\over 1+t} {{\hat f}(t,z)-1\over {\hat h}(t,z)}\cr}}
with $p_1$ as in \valqbis.

Note that, in the same way as $e_\ell$ and $g_\ell$ are the fully well-balanced
counterparts of $f_\ell$ and $h_\ell$,  $p_\ell$ may be viewed as
the fully well-balanced counterpart of $q_\ell$. Writing $f_\ell=q_\ell+
q_{\ell+1}$, we may, after some simple algebra, rewrite \eplink\ as
\eqn\qplink{{\hat p}(t,z)=p_1(z)+{{\hat q}(t,z)-1\over {\hat h}(t,z)}
\quad {\rm with}\ {\hat q}(t,z)\equiv \sum_{\ell\geq 0}q_{\ell+1}t^\ell\ .}

\newsec{Distance statistics in large quadrangulations with no multiple edges}

\subsec{Simple enumerations}
As a non-trivial check of the formulas of Section 2.4, we may verify some sum
rules by computing
${\hat e}(1,z)$, ${\hat g}(1,z)$ and ${\hat p}(1,z)$ which correspond
to quadrangulations with two marked ``points'' (vertices or edges)
without constraint on the distance. Using 
$\sum_{\ell=1}^{\ell_{\rm max}}f_\ell=r_{\ell_{\rm max}+1}+
r_{\ell_{\rm max}}-r_1$, we deduce that
\eqn\fofone{{\hat f}(1,z)=2r-1}
while, from $\sum_{\ell=1}^{\ell_{\rm max}}h_\ell=r_{\ell_{\rm max}+1}
r_{\ell_{\rm max}}$, we get
\eqn\hofone{{\hat h}(1,z)=r^2\ .}
From \hatrel\ and \hatrelbis, this leads to 
\eqn\eofone{\eqalign{{\hat e}(1,z)& =2{r-1\over r^2}=2 z r =
\sum_{n\geq 1} {2\, (3n-3)!\over (2n-1)!(n-1)!} z^n\cr
{\hat g}(1,z)&=1-{1\over r^2}= z r (r+1)=
\sum_{n\geq 1} {2\, (3n-3)!\over (2n-2)!n!} z^n\ .\cr
\cr}}
These expressions are consistent with \valqbis\ as they give
\eqn\checkone{\eqalign{ 
{{\hat e}(1,z)\vert_{z^n}\over p_1\vert_{z^n}}=n \cr 
{{\hat g}(1,z)\vert_{z^n}\over p_1\vert_{z^n}}=2n-1 \cr 
}}
where we recognize the number $n=(n+2)-2$ of vertices different
from the two extremities of the root edge in a rooted quadrangulation
and the number $(2n-1)$ of edges different from the root edge. This
is expected since in ${\hat e}(1,z)$ (respectively ${\hat g}(1,z)$),
there is no condition of distance for the marked vertex (respectively 
the second marked edge). 

Finally, from \eplink, we also have
\eqn\checkone{{{\hat p}(1,z)\vert_{z^n}\over p_1\vert_{z^n}}=1+{n\over 2}
={n+2\over 2}}
consistent with the fact that ${\hat p}(1,z)$ and $p_1$ both count
maps with a marked edge, with an extra origin vertex in ${\hat p}(1,z)$
($n+2$ choices) and an extra orientation for the marked edge 
in $p_1$ ($2$ choices).

From the explicit forms \valfl\ and \valhl\ of $f_\ell$ and $h_\ell$, 
and via the relations \ghlinkbis, \eflinkbis\ and \perel\ (or their 
compact forms \hatrel, \hatrelbis\ and \eplink), we have an implicit
access to the desired functions $e_\ell$, $g_\ell$ and $p_\ell$. 
Explicit expressions may easily be obtained for the first values of $\ell$.
Taking $\ell=1,2,3$, we deduce for instance the first terms in the $z$ 
expansion of $p_\ell$, $g_\ell$ and $e_\ell$:
\eqn\smallz{\eqalign{
p_1(z)&=2z+z^2+2z^3+6z^4+22z^5+91z^6+408z^7+1938z^8+9614z^9+\cdots\cr
p_2(z)&=g_1(z)=e_1(z)\cr 
& =z+z^2+3z^3+11z^4+46z^5+209z^6+1006z^7+5053z^8+26227z^9+\cdots\cr
g_2(z)&= 
    z+2z^2+7z^3+29z^4+132z^5+639z^6+3232z^7+16896z^8+90643z^9+ \cdots\cr 
e_2(z)&= 
    z+z^2+3z^3+12z^4+55z^5+272z^6+1411z^7+7565z^8+41560z^9+ \cdots\cr 
p_3(z)&= 
    z^4+9z^5+63z^6+405z^7+2512z^8+15333z^9+ \cdots\cr 
g_3(z)&= 
    2z^4+20z^5+151z^6+1030z^7+6705z^8+42617z^9+ \cdots\cr
e_3(z)&= 
    z^4+9z^5+64z^6+422z^7+2698z^8+17011z^9+ \cdots\cr
}}

More explicit distance-dependent results may be extracted from the
above implicit form of $e_\ell$, $g_\ell$ or $p_\ell$ {\it in the limit 
of large quadrangulations}, i.e.\ when $n\to \infty$. This can be done
by keeping $\ell$ finite, giving rise to the so-called local limit,
or by letting $\ell$ scale as $n^{1/4}$, leading to the so-called
scaling limit. Let us now discuss these two cases in details.

\subsec{Local limit laws for large quadrangulations with no multiple 
edges}
The large $n$ asymptotics of quadrangulations with no multiple edges
may be extracted from the singular behavior of the various generating 
functions above when $z$ approaches its critical value, equal to
$4/27$ as apparent for instance in \valr.
Writing 
\eqn\expanz{z={4\over 27}(1-\eta^2)}
we have for instance the small $\eta$ expansion
\eqn\smet{p_1={1\over 3}-{4\over 9} \eta^2+{8\over 27 \sqrt{3}} \eta^3+\cdots}
with no term proportional to $\eta$, so that the leading singular part 
of $p_1$ is coded by the coefficient of $\eta^3$. We immediately 
deduce the large $n$ behavior
\eqn\larn{p_1\vert_{z^n}\sim \left({27\over 4}\right)^n\, {8\over 27 \sqrt{3} 
n^{5/2}}
{1\over \Gamma(-3/2)}=\left({27\over 4}\right)^n\, {2\over 9 \sqrt{3 \pi}
n^{5/2}}}
a result which can alternatively be obtained from the explicit expression
of $p_1\vert_{z^n}$ read off \valqbis. Similarly, we get from the explicit 
forms \valfl\ 
and \valhl\ of $f_\ell$ and $h_\ell$ small $\eta$ expansions for 
${\hat f}(t,z)$ and ${\hat h}(t,z)$ of the form
\eqn\smalleta{\eqalign{
{\hat f}(t,z)&= {\hat A}^{(f)}(t) + {\hat C}^{(f)}(t)\, \eta^2 + 
{\hat D}^{(f)}(t)\, \eta^3 +\cdots\cr
{\hat h}(t,z)&= {\hat A}^{(h)}(t) + {\hat C}^{(h)}(t)\, \eta^2 + 
{\hat D}^{(h)}(t)\, \eta^3 +\cdots\cr}}
with coefficients which can be computed explicitly, and with again a 
leading singular behavior coded by the coefficient of 
$\eta^3$.  We deduce from \ghlinkbis\ and \eflinkbis\ the expansions
\eqn\smalletabis{\eqalign{
{\hat e}(t,z)&= {\hat A}^{(e)}(t) + {\hat C}^{(e)}(t)\, \eta^2 + 
{\hat D}^{(e)}(t)\, \eta^3 +\cdots\cr
{\hat g}(t,z)&= {\hat A}^{(g)}(t) + {\hat C}^{(g)}(t)\, \eta^2 + 
{\hat D}^{(g)}(t)\, \eta^3 +\cdots\cr}}
where, in particular,
\eqn\AADD{\eqalign{
{\hat D}^{(e)}(t)&={{\hat D}^{(h)}(t)-{\hat A}^{(f)}(t){\hat D}^{(h)}(t)+
{\hat A}^{(h)}(t){\hat D}^{(f)}(t)\over ({\hat A}^{(h)}(t))^2} \cr
{\hat D}^{(g)}(t)&={{\hat D}^{(h)}(t)\over ({\hat A}^{(h)}(t))^2}\ .\cr}}
Doing the explicit computation for these two quantities, we find:
\eqn\explidhatg{\eqalign{
& {\hat D}^{(e)}(t) ={4(t+1)\over 
945 \sqrt{3}\, t\, (1-t)^4 (t (3 t-4)+4 (1-t) {\rm Li}_2(t))^2}\cr
&\times \Big\{
t\Big(12(1\!-\!t)^2\log(\!1\!-\!t)\ 
(9t^5\!-\!13t^4\!-\!49t^3\!+\!128t^2\!-\!57t\!+\!12)
\cr
&\qquad \qquad \qquad \qquad +\!t(\!-\!175t^6\!+\!554t^5\!+\!89t^4\!-\!2292t^3\!+\!2610t^2
\!-900t\!+\!144)\!\Big)\cr
&\ \ -\!12(1\!-\!t) {\rm Li}_2(t)
\Big(12(\!1\!-\!t)^6 \log(\!1\!-\!t\!)
+\!t
(\!8t^6\!-\!33t^5\!+\!64t^4\!-\!103t^3\!+\!148t^2\!-\!66t\!
+\!12)\!\Big)
\!\Big\}
\cr
& {\hat D}^{(g)}(t) =
{16 t^3 \left(t \left(35 t^4-151 t^3+236 t^2-114 t+24\right)
-24 (1-t)^5{\rm Li}_2(t) \right)\over 945 \sqrt{3}\, (1-t)^4 
\left(t (3 t-4)+4 (1-t) 
{\rm Li}_2(t)\right)^2}
\cr}}
where ${\rm Li}_2(t)=\sum_{k\geq 1}{t^k\over k^2}$.
These functions encode the average number $\langle V_\ell\rangle$ of vertices
at distance $\ell$ (respectively $\langle E_\ell\rangle$ of 
edges of type $(\ell-1)\to\ell$) in rooted quadrangulations with no 
multiple edges in the limit of large size $n$. Indeed, we have 
\eqn\avene{
\eqalign{
\langle V_\ell \rangle &  = \delta_{\ell,1}+\lim_{n\to \infty}
{e_\ell\vert_{z^n}\over p_1\vert_{z^n}} = \delta_{\ell,1}
+{27 \sqrt{3}\over 8} {\hat D}^{(e)}\vert_{t^{\ell-1}}\cr
\langle E_\ell \rangle &  = \delta_{\ell,1}+\lim_{n\to \infty}
{h_\ell\vert_{z^n}\over p_1\vert_{z^n}} = \delta_{\ell,1}
+{27 \sqrt{3}\over 8} {\hat D}^{(g)}\vert_{t^{\ell-1}}\cr
}}
or equivalently
\eqn\avenebis{\eqalign{
\sum_{\ell\geq 0} t^\ell \langle V_{\ell+1} \rangle & =
1+{27 \sqrt{3}\over 8}{\hat D}^{(e)}(t)\cr
\sum_{\ell\geq 0} t^\ell \langle E_{\ell+1} \rangle & =
1+{27 \sqrt{3}\over 8}{\hat D}^{(g)}(t)\ .\cr
}}
Expanding ${\hat D}^{(e)}(t)$ and ${\hat D}^{(g)}(t)$ 
at small $t$, we get for instance
\eqn\firstVEl{\eqalign{
\langle V_1 \rangle & = {133\over 25} \ , \quad
\langle V_2 \rangle = {1809\over 125}\ , \quad
\langle V_3 \rangle = {90747\over 3125} \cr
\langle E_1 \rangle & = {133\over 25} \ , \quad
\langle E_2 \rangle = {2727\over 125}\ , \quad
\langle E_3 \rangle = {598563\over 12500} \ .\cr
}}
For $t\to 1$, we have 
\eqn\tonedge{{\hat D}^{(e)}(t) \sim {16\over 63 \sqrt{3}} {1\over 
(1-t)^4}\ ,\quad
{\hat D}^{(g)}(t) \sim {32\over 63 \sqrt{3}} {1\over 
(1-t)^4}}
from which we deduce the large $\ell$ behaviors
\eqn\largelbe{
\langle V_\ell \rangle \sim {1\over 7}\ell^3\ , \quad
\langle E_\ell \rangle \sim {2\over 7}\ell^3\ .}
We may finally extract from the expansion of ${\hat p}(t,z)$ the average
number $\langle E_\ell \rangle_\bullet$ of edges at distance
$\ell$ from the origin now in the ensemble of pointed quadrangulations
of fixed large size $n$.
We find 
\eqn\firstElpointed{\eqalign{
\langle E_1 \rangle_\bullet & = 4 \ , \quad
\langle E_2 \rangle_\bullet = {432\over 25}\ , \quad
\langle E_3 \rangle_\bullet = {5076\over 125} \cr
}}
i.e.\ the statistics of neighbors are slightly different in rooted 
and pointed quadrangulations, as expected. Still, for large $\ell$, we
get
\eqn\largelpoint{
\langle E_\ell \rangle_\bullet \sim {2\over 7}\ell^3}
and pointed and rooted maps give rise to the same statistics in this limit.

\subsec{Scaling limit laws for large quadrangulations with no multiple 
edges}
A non-trivial scaling limit is obtained by letting $z$ tend to its
critical value $4/27$ and letting $\ell$ become large as
\eqn\scallim{z={4\over 27}(1-\eta^2)\ ,\qquad \ell=L \eta^{-1/2}}
with $\eta\to 0$ and $L$ finite. 
Using the small $\eta$ behavior $y\sim e^{-2 \alpha\, 
\eta^{1/2}}$ with $\alpha=3^{1/4}/\sqrt{2}$, we find the leading behavior
\eqn\leadhell{h_\ell\sim -{9\over 2}\, \eta^{3/2}\ {\cal F}'\left(
L;3^{1/4}/\sqrt{2}\right)}
where we introduce the scaling function  
\eqn\calf{{\cal F}(L;\alpha)={2\alpha^2\over 3}
\left(1+{3\over \sinh(\alpha L)^2}\right)}
and where ${\cal F}'$ stands for the derivative of ${\cal F}$ with 
respect to $L$. We recognize here the scaling function ${\cal F}$ encountered
\GEOD\ when computing the scaling limit of the two-point function in the 
context of general (with possibly multiple edges) quadrangulations.
Note that the value of $\alpha$ in this latter case is however different, 
equal to $\sqrt{3/2}$. This is not a surprise since, from the substitution
approach, the vicinity of the critical point $z=4/27$ corresponds
precisely to the vicinity of the 
critical point $g=1/12$ for general quadrangulations. 
From \zgz, we indeed have the expansion
\eqn\zgzscal{g\left({4\over 27}(1-\eta^2)\right)={1\over 12}\left(1-
{1\over 3} \eta^2+
{\cal O}(\eta^3)\right)}
so that expanding in $\sqrt{1-27z/4}$ amounts to expanding in 
$\sqrt{3}\, \sqrt{1-12g}$ and the change of normalization by $\sqrt{3}$ simply
translates into a change of the value of $\alpha$ by $3^{1/4}$.

Rather than that of $h_\ell$, let us now explore the scaling form of $g_\ell$ 
itself, which is relevant to the statistics of distances in quadrangulations
with no multiple edges. This requires a few manipulations as follows.
Right at the critical value $z=4/27$, we have
\eqn\hlcrit{h_\ell(4/27)= 9 {(2\ell+3)\over (\ell+1)^2(\ell+2)^2}}
so that, in the scaling limit \scallim, we may write at leading order in $\eta$ 
\eqn\leadhellbis{h_\ell(z)-h_\ell(4/27) \sim \eta^{3/2}
\left( -{9\over 2}\ {\cal F}'\left(L;3^{1/4}/\sqrt{2}\right) 
-{18\over L^3}\right)\ .}
Setting 
\eqn\scalt{t=1-s \eta^{1/2}}
this translates into
\eqn\scalht{{\hat h}(t,z)-{\hat h}(t,4/27)\sim\ 
\eta\  \int_0^\infty dL\ e^{-s\, L} 
\left(-{9\over 2}\ {\cal F}'\left( L;3^{1/4}/\sqrt{2}\right)
-{18\over L^3} \right)}
with, from \hlcrit, 
\eqn\hcritt{\eqalign{{\hat h}(t,4/27) & =-{9 
\left(t (3 t-4)+4 (1-t) {\rm Li}_2(t)\right)\over 4 t^3}\cr
 &= {9\over 4}\!+\!{3\over 4} (15\!-\!2\pi^2)\, s\eta^{1/2}\!-\!{9\over 4}
\left(s^2 \, \eta \left(4 \log (s\, \eta^{1/2})\!+\!2 \pi^2\!-\!13
\right)\right)\!+\!{\cal O}\left(\eta^{3/2}\right)\ . \cr}}
From \hatrel, this leads to 
\eqn\scalgt{{\hat g}(t,z)-{\hat g}(t,4/27) \sim\ 
{16\over 81} \ \eta\ \int_0^\infty dL\ e^{-s\, L}
\left(-{9\over 2}\ {\cal F}'\left( L;3^{1/4}/\sqrt{2}\right)
-{18\over L^3} \right)}
with
\eqn\gcritt{\eqalign{{\hat g}(t,4/27)& = 
1-{1\over {\hat h}(t,4/27)} \cr
 &= {5\over 9}\!+\!{4\over 27} (15\!-\!2\pi^2)\, s \eta^{1/2}\!-\!{8\over 81}
\left(s^2 \eta\, \left(18 \log (s \eta^{1/2})\!+\!2\pi^4\!-\!21\pi^2\!+\!54
\right)\right)\! 
+\!{\cal O}\left(\eta^{3/2}\right) \cr}}
From the singular part of ${\hat g}(t,4/27)$ when $t\to 1$ (i.e.\
$s\to 0$), we deduce the large $\ell$ behavior
\eqn\largeellg{g_\ell(4/27)\sim {32\over 9 \ell^3}=\eta^{3/2}\,
{32\over 9 L^3}}
and from \scalgt, we deduce eventually
\eqn\scalgell{\eqalign{& 
g_\ell-g_\ell(4/27)\sim \eta^{3/2}
\left(-{8\over 9}\ {\cal F}'\left( L;3^{1/4}/\sqrt{2}\right) 
- {32 \over 9 L^3}\right)\cr
&{\rm i.e.}\ \ 
g_\ell\sim -{8\over 9}\, \eta^{3/2}\, 
{\cal F}'\left( L;3^{1/4}/\sqrt{2}\right)\ .\cr}
}
The net result of passing from $h_\ell$ to $g_\ell$ is therefore a simple
normalization factor ($8/9$ for $g_\ell$ instead of $9/2$ for $h_\ell$).
This factor is important as it will ensure a proper normalization of the 
two-point function but $g_\ell$ behaves essentially as $h_\ell$ which,
from the substitution approach, is characterized by the same scaling function
as for general quadrangulations, up to a change of scale by $3^{1/4}$.

Now the above result may be used to study the two-point distance statistics
in the ensemble of quadrangulations with no multiple edges having
a fixed size $n$ ($=$ number of faces), in the limit of large
$n$ . As explained in \GEOD,
the term of order $z^n$ in the various generating functions above
may be extracted by a contour integral in $z$ which at large
$n$, translates via a saddle point estimate into an integral
over a real variable $\zeta$ upon setting
\eqn\scalzzeta{z={4\over 27}\left(1+{\zeta^2\over n}\right)\ .}
Setting now
\eqn\scalln{\ell=r n^{1/4}} we may use the above formulas
with $\eta=-{\rm i}\zeta\, n^{-1/2}$ and $L=\sqrt{-{\rm i}\zeta}\ r$ 
to deduce the probability ${\tilde \rho}(r)$ that a marked edge picked uniformly
at random in a rooted quadrangulation with no multiple
edges be at rescaled distance $r$ from the root edge
\eqn\rhoofr{\eqalign{{\tilde \rho}(r)&=\lim_{n\to \infty} n^{1/4}\ {1\over 2n}\ 
{g_\ell\vert_{z^n}\over p_1\vert_{z^n}}\cr
&= \lim_{n\to \infty} n^{1/4}{1\over 2n} 
{9 \sqrt{3 \pi} \over 2} n^{5/2} {1\over \pi\, n}
\int_{-\infty}^{\infty}d\zeta\, (-{\rm i} \zeta)e^{-\zeta^2}
\left({-{\rm i}\zeta\over n^{1/2}}\right)^{3/2}
\left(-{8\over 9}\right) 
{\cal F}'\left( \sqrt{-{\rm i}\zeta}\ r;3^{1/4}/\sqrt{2}\right)\cr
&= {2 \sqrt{3}\over \sqrt{\pi}} \int_{-\infty}^{\infty}d\zeta 
\, {\rm i} \zeta\ e^{-\zeta^2}
{\cal F}'\left(r;3^{1/4}/\sqrt{2}\sqrt{-{\rm i}\zeta}\right)
\cr}
}
where we used the self-similarity property 
${\cal F}'(u\, r;\alpha)= u^{-3} {\cal F}'(r;\alpha\, u)$.

We could repeat all the above calculations for $e_\ell$ instead of $g_\ell$.
This leads to the same scaling form ${\tilde \rho}(r)$ for the probability
that a marked vertex picked uniformly at random in a rooted quadrangulation 
with no multiple edges be at rescaled distance $r$ from the root edge.
This is expected as vertices cannot be distinguished from edges in
the scaling limit. We could also use $p_\ell$ and again, we would get 
the same scaling form for the probability that a marked edge picked uniformly 
at random in a pointed quadrangulation with no multiple edges be at rescaled 
distance $r$ from the root origin vertex.

We may now compare the above result with the density of points 
(vertices or edges) at distance $\ell=r n^{1/4}$ in general quadrangulations of
size $n$, in the limit $n\to \infty$. It is given by \GEOD 
\eqn\rhohab{\rho(r)={2 \over \sqrt{\pi}} \int_{-\infty}^{\infty}d\zeta 
\, {\rm i} \zeta\ e^{-\zeta^2}
{\cal F}'\left(r;\sqrt{3/2}\sqrt{-{\rm i}\zeta}\right)
}
so that, using again the self-similarity of ${\cal F}'$, we may write 
\eqn\rhorhotilde{{\tilde \rho}(r)={1\over 3^{1/4}} \rho\left({r\over 3^{1/4}}
\right)\ .}
This may be viewed as a manifestation of the universal nature of the two-point
function, which is the same for the two classes (with or without multiple
edges) of maps, up to some global (non-universal) scale. 
In the present case, this universality can simply be traced back in the 
calculations and comes from the fact that passing from $H_\ell$ to
$h_\ell$ is a simple substitution $g\leftrightarrow z$ (which explains the 
factor $3^{1/4}$) and then passing from $h_\ell$ to $g_\ell$ simply
amounts to some global normalization. 

Since the rescaled distance is 
obtained by normalizing the original distance
by $n^{1/4}$, the rescaling factor above states that general quadrangulations 
with size $n$ are asymptotically isometric to quadrangulations with no 
multiple edges of size $(n/3)$.
This result is consistent with the heuristic picture of a general 
quadrangulation as made of a large component, the {\it ``mother universe''}, 
with no multiple edges and of typical size $n/3$, and of small components, 
the {\it ``baby universes''}, of finite extension. 
The two-point function of general quadrangulations is entirely dictated
by the distance statistics in their mother universes. This is
because the parts of geodesics lying in baby universes are of 
negligible length. 

\newsec{Mother universe and minbus in general quadrangulations}

In this Section, we return to the case of general quadrangulations
and discuss their geometry in the light of our new results. As just
explained, the geometry of the mother universe itself is well captured 
by the two-point function. On the contrary, this two point-function is
blind to the geometry of minimal neck baby universes. Here we compute a 
number of distance-dependent quantities characterizing the minbus. 
To this end, let us first discuss how our various generating functions 
for almost well-balanced trees translate in the language of general 
quadrangulations. 
Their meaning is best captured by reformulating the substitution of Section
2.3 directly {\it at the level of the maps}.

\subsec{Substitution for rooted maps}

Out of any {\it rooted} general quadrangulation, we may extract 
a quadrangulation with no multiple edges as follows. Consider
two fixed vertices adjacent in the quadrangulation and connected by
exactly $k$ edges ($k\geq 1$), which we may orient for convenience from the
vertex closest to the origin of the root edge to the other vertex. 
These edges separate the map into $k$ {\it elementary domains} 
and the {\it root face} of the map, i.e.\ the face immediately
on the left of the root edge, belongs to exactly one of them.
Let us consider the ``outgrowth'' formed by the union of the 
$(k-1)$ elementary domains not containing the root face. 
This outgrowth is either empty (if $k=1$) or it is a quadrangulation 
with a 
boundary of length $2$ which, upon continuous deformation in the sphere, 
may be glued into a single oriented edge. The outgrowth may therefore 
be described as a (possibly empty) rooted quadrangulation itself. Moreover, 
if we now consider all pairs of vertices linked by more than one edge,
and the corresponding (non-empty) outgrowths, the interiors
of these outgrowths form a forest 
(i.e.\ any two of them are either disjoint
or included one into the other), which allows us to select {\it maximal 
outgrowth} (i.e.\ those not contained in a bigger one). 
If we now squeeze into a 
single edge each of these maximal outgrowths, we end up with a 
quadrangulation 
with no multiple edges which is naturally rooted at the original root edge. 
Conversely, a general rooted quadrangulation is obtained from a 
quadrangulation with no multiple edges by inflating each edge into an
outgrowth as above. 
If we assign a weight $g$ per face of the general quadrangulation, 
the generating function describing a squeezed outgrowths is that, $R_1(g)$, of
rooted quadrangulations (including a first term $1$ in $R_1$ for the empty 
case). 
This results in the identity
\eqn\roneqlink{R_1(g)=1+p_1\left(g\, \left(R_1(g)\right)^2\right)}
relating the generating function $R_1$ of rooted general quadrangulations
to that $p_1$ of rooted quadrangulations with no multiple edges.
In \roneqlink, we used the fact that there are twice as many edges 
as faces in a quadrangulations, so assigning a weight $g$ per face
and a weight $R_1(g)$ per edge is equivalent to 
assigning a weight $z=g\ (R_1(g))^2$ per face only. Introducing 
$z(g)$ as in \zgz, and its inverse $g(z)$, equation \roneqlink\ reads
$p_1(z)=R_1(g(z))-1$. Using $R_1=R-g R^3=(1+3g R^2)(1-g R^2)$ at $g=g(z)$
(where $g R^2=z r^2$), we recover $p_1=(1+3 z r^2)(1-z r^2)-1=z r (3-r)$
as expected. 
The above construction allows us 
to define for each rooted quadrangulation its {\it core}, which is
the rooted quadrangulations with no multiple edges obtained after the squeezing
procedure.

The above connection was studied in details in Ref. \BaFlScSo\ in the slightly 
different, 
but equivalent context of general and $2$-connected maps. Translated into
our language, the result is that, in the ensemble of 
rooted quadrangulations with a fixed 
size $n$, and in the limit $n\to\infty$, the law for the size 
(number of faces) of the core is bimodal: with probability $2/3$, the core 
remains finite while, with probability $1/3$, it is macroscopic, of size $n/3
+ \alpha n^{2/3}$ with $\alpha$ distributed according to a {\it standard
Airy distribution}. Rather than focusing on the core itself, which 
depends strongly on the choice of root edge, one may fully decompose the 
rooted quadrangulation into components with no multiple edges by applying 
recursively the squeezing procedure inside each of the squeezed outgrowths 
(which may themselves be viewed as rooted quadrangulations as explained above).
Alternatively, this procedure amounts to disconnect the map into
pieces by cutting along all its minimal necks, which are the cycles of 
length $2$ in the map and then gluing the necks into single edges. Each of the 
obtained component is a rooted quadrangulation with no multiple edges. 
In the limit of large $n$, the largest component has size $n/3
+\alpha n^{2/3}$ as above while the second largest one has size at most 
$n^{2/3}$. 
Adopting the terminology of quantum gravity, the largest component 
defines precisely the mother universe, while the other components,
once reglued together, form the minimal neck baby universes (minbus). 
A detailed heuristic 
discussion of minbus can be found in Ref.\JainMa. In particular, it
is shown that, in all generality, the largest minbu has a size of order 
$n^{1/(1-\gamma_{\rm string})}$ where $\gamma_{\rm string}$ is the so-called 
string susceptibility exponent, equal to $-1/2$ for pure gravity.

\subsec{Substitution for maps with two marked points}
\fig{The squeezing procedure for a rooted map with a marked vertex at
distance $\ell$ (distinct from the endpoint of the root edge if $\ell=1$). 
If the root edge itself is a multiple edge, we squeeze all the associated 
elementary domains (filled regions) 
except that containing the marked vertex. For any other pair of vertices
linked by multiple edges, we squeeze the associated elementary domains 
(filled regions) except that containing the root edge and that containing 
(strictly) the marked vertex. These two unsqueezed domains may be the same
or not. In practice, it is sufficient to squeeze the maximal domains 
only.}{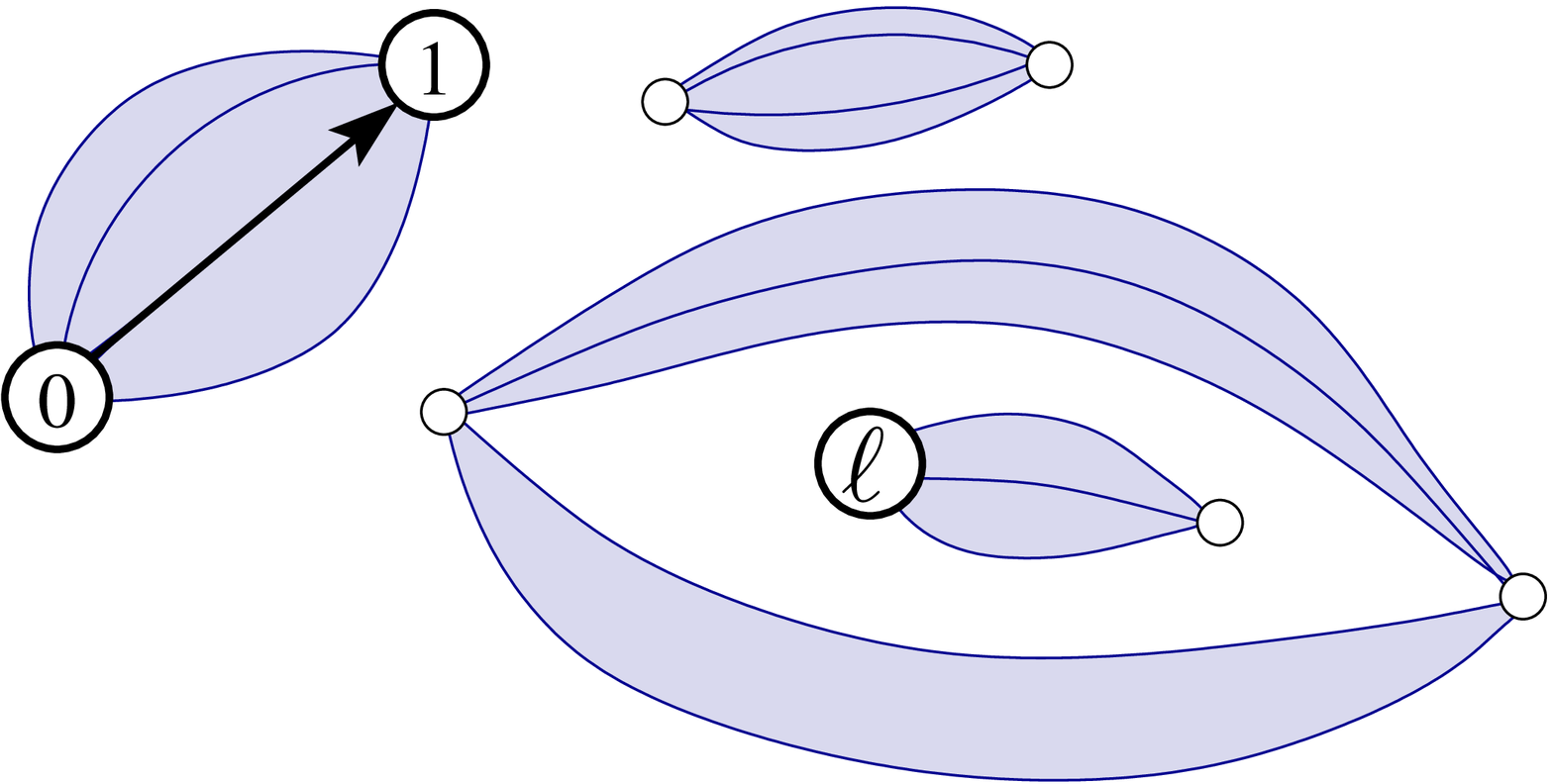}{8.cm}
\figlabel\Fellsqueez
The above squeezing procedure dealt with quadrangulations with a marked root 
edge only, as counted by $R_1$. We now wish to apply a similar squeezing 
procedure to the configurations with two marked points (edge or vertex)
distant by $\ell$. More precisely, let us start with general rooted 
quadrangulations with an extra marked vertex 
at distance $\ell$ from the origin of the root edge, as counted by $F_\ell$. 
Let us suppose first that $\ell>1$ so that, in particular, 
the marked vertex is different from the endpoint of the root edge.
If the extremities of the root edge are 
linked by $k>1$ edges, delimiting $k$ elementary domains, we start by 
squeezing into a single oriented edge the outgrowth formed by the union of 
the $(k-1)$ such elementary domains not containing the marked vertex
(see figure \Fellsqueez\ for an illustration). This oriented edge becomes 
the new root edge. Note that the choice of outgrowth here is slightly different 
from that adopted above for rooted quadrangulations without a marked vertex. 
In particular the outgrowth is now formed in general of two pieces, 
one on each side of the original 
root edge, and separated on the sphere by the marked vertex. 
Its contribution to
$F_\ell$ is thus counted by $(R_1(g))^2$. For any other 
pair of vertices, we look at the 
outgrowth formed by the union of all the associated elementary 
domains except that containing the root edge and, if the marked vertex 
is not one of the two vertices in the pair, that containing 
the marked vertex. Note that this second excluded elementary domain 
may be the same as the first one or not, so that
the outgrowth consists either of one connected piece or of two pieces
separating the root edge from the marked vertex
(see figure \Fellsqueez). Considering
all pairs of vertices linked by more than one edge, the interior of
the corresponding (non-empty) outgrowths form again a forest among which we 
select the maximal outgrowths that we squeeze into single edges. After
squeezing, we end up with a rooted quadrangulation with an extra marked 
vertex at distance $\ell$ from the origin of the root edge, and satisfying
\item{$\bullet$ (1)}{all edges are simple or double edges.} 
\item{$\bullet$ (2)}{the root edge is a simple edge.}
\item{$\bullet$ (3)}{a double edge necessarily separates strictly 
the root edge from the marked vertex.}
\par
\noindent The fact that the marked vertex is still at distance 
$\ell$ from the origin of the root edge is
because the original geodesic paths never enter strictly the outgrowths which 
have been squeezed, so these paths are still present in the squeezed
object and no shorter path has been created.  If we denote by
$f_\ell(z)$ the generating function for such quadrangulations with 
a weight $z$ per face, the above substitution procedure translates into the 
relation
\eqn\newrelfF{F_\ell(g)=R_1(g) \ 
f_\ell\left(g \left(R_1(g)\right)^2\right)}
since each edge of the quadrangulations counted by $f_\ell$ has to be
substituted by an outgrowth (counted by $R_1(g)$), 
except for the root edge whose outgrowth is counted by $(R_1(g))^2$, 
hence the extra multiplicative factor $R_1(g)$. The above
relation precisely matches \substFH, so our new definition of $f_\ell$
matches its former definition of Section 2.4 in terms of chains of trees.
In other words, we end up with a direct interpretation of the quantity 
$f_\ell$ of
Section 2.4 as counting rooted quadrangulations with a marked vertex at 
distance $\ell$, 
and satisfying (1)-(3) above. 
\fig{The generating function $f_\ell$ enumerates rooted quadrangulations
with a marked vertex at distance $\ell$ from the origin of the root edge
(and different from its endpoint if $\ell=1$), with only single and
double edges. The root edge is necessarily a single edge and double 
edges necessarily separate (strictly) the marked vertex from the root edge.
For $\ell=1$, a term $1$ is added to $f_\ell$ to guarantee relation
\newrelfF\ or equivalently \substFH.}{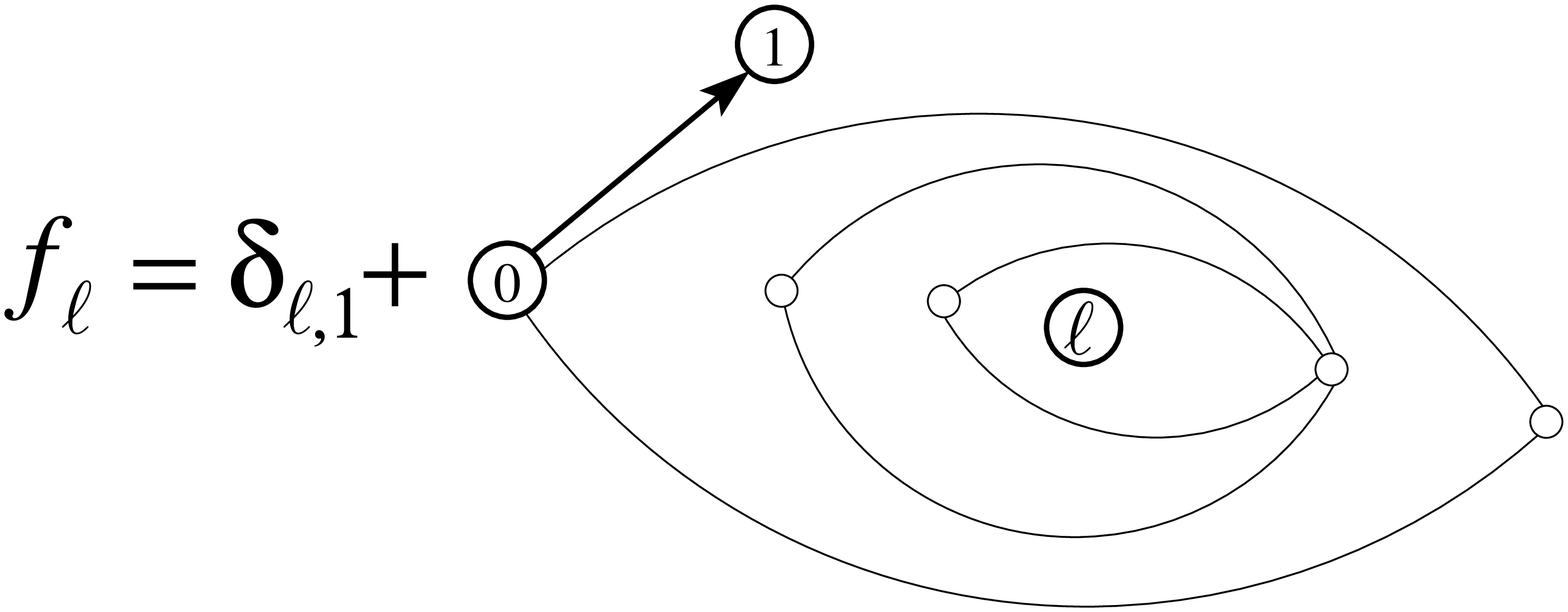}{8.cm}
\figlabel\fell
If $\ell=1$, i.e.\ if we start with a rooted quadrangulation with an extra
marked vertex at distance $1$ from the origin, we may apply exactly
the same squeezing procedure provided that the marked vertex be different
from the endpoint of the root edge. If the marked vertex
happens to be the endpoint of the root edge, we decide to squeeze all the 
associated elementary domains, resulting into a trivial object made of the 
root edge itself. This procedure ensures that equation \newrelfF\ above holds 
for $\ell=1$: $f_1(z)$ contains a constant term $1$ accounting for the
trivial case above while $f_1(z)-1$ is the generating function for rooted 
quadrangulations with an extra marked vertex adjacent to the origin 
but non-incident to the root, and satisfying (1)-(3) above.
The interpretation of $f_\ell$ for $\ell\geq 1$ is illustrated in figure \fell.
\fig{The squeezing procedure for a rooted map with a marked edge of
type $(\ell-1)\to\ell$ with respect to the origin of the root edge (and
whose endpoint is distinct from the endpoint of the 
root edge if $\ell=1$). If the root edge itself is a multiple
edge, we squeeze all the associated elementary domains (filled regions) 
except that containing the marked vertex. Similarly, if the marked edge is a 
multiple edge, we squeeze all the associated elementary domains (filled
regions) except that containing the root edge. For any other pair of vertices
linked by multiple edges, we squeeze the associated elementary domains 
(filled regions) except that containing the root edge and that containing 
the marked edge. These two unsqueezed domains may be the same
or not. In practice, it is sufficient to squeeze the maximal 
domains.}{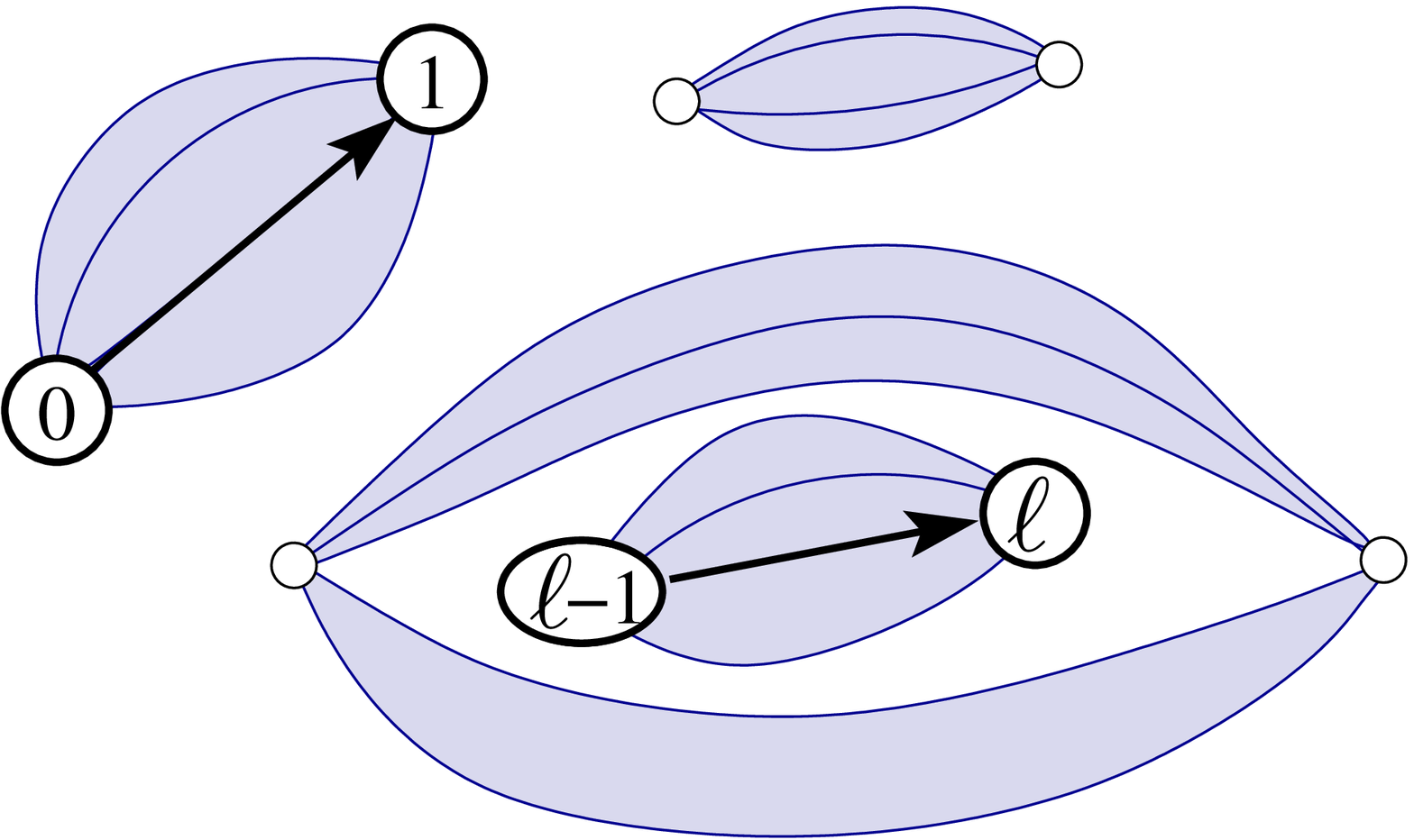}{8.cm}
\figlabel\Hellsqueez

\fig{The generating function $h_\ell$ enumerates rooted quadrangulations
with a marked edge of type $(\ell-1)\to\ell$ with respect to the origin of 
the root edge (and with its endpoint different from the endpoint 
of the root edge if $\ell=1$), with only single and
double edges. The root edge and the marked edge are necessarily single edges 
and double edges necessarily separate the marked edge from the root edge.
For $\ell=1$, a term $1$ is added to $h_\ell$ to guarantee relation
\substFH.}{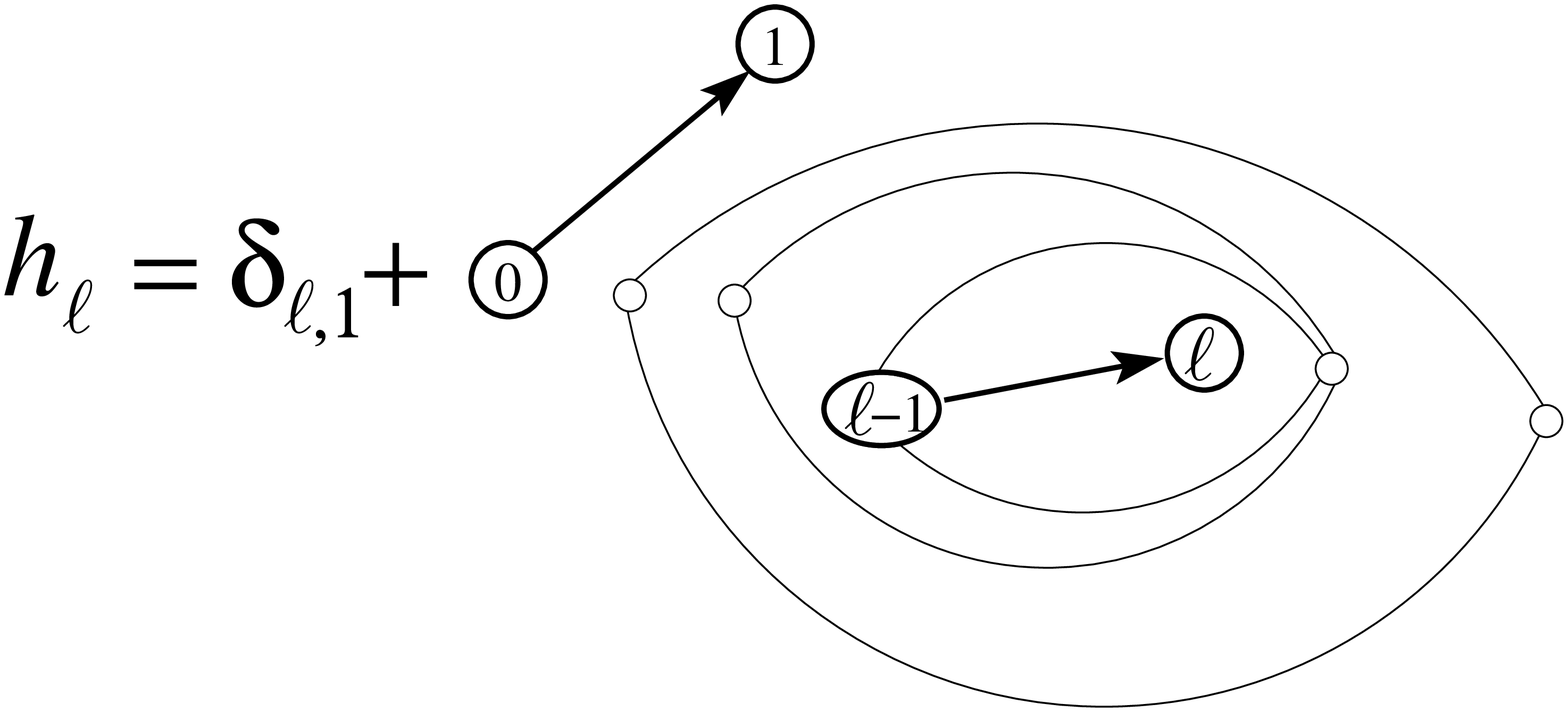}{8.cm}
\figlabel\hell

As for $H_\ell$, recall that it counts rooted general 
quadrangulations with an extra marked edge of type $(\ell-1)\to\ell$ with 
respect to the origin of the root edge. Assuming again that $\ell>1$ so that 
this marked edge has an endpoint different from that of the root edge, 
we squeeze the outgrowths on both sides
of the root edge and on both sides of the marked edge, transforming
these edges into single edges (see figure \Hellsqueez). Repeating the above 
arguments, we see that, upon squeezing of the maximal outgrowth not containing 
the two marked edges, $H_\ell$ is related via \substFH\ to the generating 
function $h_\ell(z)$ for rooted 
quadrangulations with an extra marked edge of type $(\ell-1)\to\ell$
with respect to the origin of the root edge, and satisfying
\item{$\bullet$ (1)'}}{all edges are simple or double edges.} 
\item{$\bullet$ (2)'}{the root edge and the extra marked edge are simple edges.}
\item{$\bullet$ (3)'}{a double edge necessarily separates strictly 
the root edge from the extra marked edge}
\par
\noindent Again when $\ell=1$, Eq. \substFH\ still holds with $h_1$ composed
of a trivial constant term $1$ (accounting for the case where the root edge 
and the extra marked edge have the same endpoint in the configuration counted 
by $H_1$) and of the generating function $h_1-1$ of rooted quadrangulations
with an extra marked edge incident to the origin of the root edge 
but with a different endpoint, and satisfying (1)'-(3)'.
This interpretation of $h_\ell$ is illustrated in figure \hell.

We may now easily understand the relations \ghlink\ and \eflink\ 
directly in the language of quadrangulations.  
From (3) or (3)', the double edges in $f_\ell$ or $h_\ell$ are necessarily 
{\it nested} and of type $(k-1)\to k$ for increasing (non-necessarily 
consecutive) values of $k$ in the range $[1,\ell]$, if we sort them
from the root edge to the marked vertex (respectively edge) at distance $\ell$. 
Moreover, the marked vertex 
(respectively the endpoint of the marked edge) at distance $\ell$ 
from the origin of the map is at distance $\ell+1-k$ of the first
extremity (i.e.\ the vertex at distance $k-1$ from the origin of the map) 
of any double edge of type $(k-1)\to k$. This is because any geodesic path
from the origin to the marked vertex (respectively endpoint) must
pass through one of the extremities of the double edge and we can always 
find one geodesic passing through the first extremity with label $(k-1)$.

Starting now from a configuration counted by $f_\ell$ (respectively $h_\ell$),
if it contains no double edge, it is then a configuration counted
by $e_\ell$ (respectively $g_\ell$). Otherwise,
looking at the double edge closest to the root-edge, i.e.\ corresponding
to the smallest value of $k$ above, this double edge separates the map
into two domains, one containing the root edge and the other the
marked vertex (respectively edge) at distance $\ell$. Upon squeezing this latter
domain into a simple edge of type $(k-1)\to k$, we end up with 
a rooted quadrangulation with no multiple edges and with a marked edge
of type $(k-1)\to k$, hence counted by $g_k$. The squeezed domain 
is a quadrangulation with a boundary of length $2$ which, upon continuous
deformation in the sphere, can be glued into a single oriented edge 
which defines a new root edge. This gives rise again to a rooted 
quadrangulation with now a marked vertex (respectively edge) at distance 
$(\ell+1-k)$ (respectively of type $(\ell-k)\to (\ell+1-k)$) with respect
to the origin of the new root edge. The squeezed domain satisfies
conditions (1)-(3) (respectively (1)'-(3)') and the corresponding 
generating function is therefore $f_{\ell+1-k}$ (respectively $h_{\ell+1-k}$)
for $k<\ell$ and $(f_1-1)$ (respectively $h_1-1$) if $k=\ell$.
This leads to the desired relations \eflink\ and \ghlink\ for $\ell>1$.
As before the case $\ell=1$ requires more care but it is easily seen to 
reproduce the relations of previous section.

Finally, from $f_\ell=r_{\ell+1}-r_{\ell-1}=q_{\ell}+q_{\ell+1}$, 
we have, via a simple re-rooting, an interpretation of $q_\ell$ 
for $\ell>1$ as counting pointed 
quadrangulations with a marked edge of type $(\ell-1)\to \ell$ satisfying
\item{$\bullet$ (1)''}{all edges are simple or double edges.} 
\item{$\bullet$ (2)''}{the marked edge is a simple edge.}
\item{$\bullet$ (3)''}{a double edge necessarily separates strictly 
the origin from the marked edge.}

\subsec{Geometry of minbus}
We may use the above interpretation of our generating functions 
to explore the geometry of minbus. We consider in this section 
the ensemble of {\it bi-rooted} 
quadrangulations where the two marked edges are at distance $\ell$,
as enumerated by $H_\ell$.
For any configuration in this ensemble, the squeezing procedure of
Section 4.2
produces a {\it bi-rooted} configuration with again its two marked edges 
at distance $\ell$, and satisfying (1)'-(3)' above. We shall call this
configuration the {\it kernel} of the bi-rooted quadrangulation. Note
that this is a notion slightly different from the core as it involves
two marked edges rather than one. In particular,
the kernel may now contain double edges which correspond to
minimal necks encountered along any geodesic path linking the origins
of the two marked edges. If we decompose as before the quadrangulation into
components with no multiple edge, the kernel may then be viewed as a linear 
sequence of such components glued via minimal necks. 
We may now condition the size of the original quadrangulation to be $n$
and compute the probability that its kernel has size ($=$ number of faces) $k$.
It simply reads
\eqn\pknl{{\cal P}_\ell^{(n)}(k)= {h_\ell\vert_{z^k}\times \left.R_1^2\, 
\left( g\, R_1^2\right)^k\right\vert_{g^n}
\over H_\ell\vert_{g^n}}}
and, at large $n$, the leading behavior of the $g^n$ coefficients may
be extracted by a simple singularity analysis. Setting 
$g={1\over 12}(1-\epsilon^2)$, we have
\eqn\Hellexpan{\eqalign{& H_\ell=A_\ell+C_\ell
\epsilon^2 +D_\ell \epsilon^3+\cdots\cr
& {\rm with} \ D_\ell=
{8\ell(\ell+3)(2\ell+3)(5\ell^4+30\ell^3+67\ell^2+66\ell+28)\over 
35(\ell+1)^2(\ell+2)^2}\cr}}
while 
\eqn\expancomp{R_1^2\,
\left(g\, R_1^2\right)^k=
{16\over 9}\left({4\over 27}\right)^k \left(1-(2+3k) \epsilon^2+4(1+k)
\epsilon^3 +\cdots\right)}
Picking the singular ($\propto \epsilon^3$) term, we get when $n\to \infty$
\eqn\pkinfl{
{\cal P}_\ell^{(\infty)}(k)= {64\over 9 D_\ell} 
h_\ell\vert_{z^k} (1+k)\left({4\over 27}\right)^k 
\ .}
For $\ell=1$, which corresponds to two marked edges having the same 
origin, we have in particular a probability ${\cal P}_1^{(\infty)}(0)=2/7$ 
that the kernel be empty, which happens when the marked edges also have 
the same endpoint. More interestingly, if we sum 
${\cal P}_\ell^{(\infty)}(k)$ over all (finite) values 
of $k$, we get the probability that the kernel remains finite:
\eqn\probcore{\eqalign{\sum_{k=0}^\infty
{\cal P}_\ell^{(\infty)}(k)&={64\over 9 D_\ell}
\left(h_\ell\left({4\over 27}\right)+{4\over 27}
h_\ell'\left({4\over 27}\right)\right) \cr
&={14(\ell^2+3\ell+4) \over 5\ell^4+30\ell^3+67\ell^2+66\ell+28}
\buildrel {\ell \gg 1}\over \sim {14\over 5\ell^2}
\ .\cr}
}
This is the probability that the kernel does not include the mother universe 
as one of its components, i.e.\  the probability
that we may go from one edge to the other without entering the
mother universe. In other words, this is nothing but the {\it probability
that the two marked edges lie in the same baby universe}. This 
probability tends to $0$ at large $\ell$ like $1/\ell^2$. 

By a slight change of normalization in the above calculation, we may return
to the ensemble of rooted quadrangulations of size $n$ and compute the 
{\it average number} of edges lying at distance $\ell$ from the root edge and 
defining with this root edge a kernel of size $k$. This number
is given by 
\eqn\Nknl{\eqalign{{\cal N}_\ell^{(n)}(k)&= {h_\ell\vert_{z^k}\times \left. R_1^2\, 
\left( g\, R_1^2\right)^k\right\vert_{g^n}
\over R_1\vert_{g^n}}\cr
& \buildrel {n\to \infty}\over \sim 
{8\over 3} h_\ell\vert_{z^k} (1+k) \left({4\over 27}\right)^k \ .\cr
}}
Upon summing over all (finite) values of $k$, we get 
\eqn\sumNknl{\eqalign{\sum_{k=0}^\infty {\cal N}_\ell^{(\infty)}(k)& ={8\over 3}
\left(h_\ell\left({4\over 27}\right)+{4\over 27}
h_\ell'\left({4\over 27}\right)\right)\cr & =
{6\ell(\ell+3)(2\ell+3)(\ell^2+3\ell+4)\over 5(\ell+1)^2(\ell+2)^2}
\buildrel {\ell\gg 1}\over \sim {12 \over 5}\ell\cr}}
which gives the average number of edges at distance $\ell$ from the root edge
and {\it lying in the same baby universe} as this root edge. 
Indeed keeping $k$ finite precisely amounts to conditioning the counted edge to
lie in the same baby universe as the root edge. The above quantity may
in this sense be viewed as the {\it two-point function inside a minbu}, 
which grows like $\ell$ at large $\ell$ instead of $\ell^3$ for the
local limit of the complete two-point function. Note that, even though 
the total size $n$ is fixed, the size of the considered minbu 
(that containing the root edge) is here a fluctuating quantity so that 
our two-point function inside a minbu corresponds in practice to a 
grand-canonical ensemble of minbus (technically this explains
why it involves the regular part of $h_\ell$ rather than its singular 
part).

Returning to the ensemble of quadrangulations with
two marked edges at distance $\ell$
and conditioning again the size $n$, we may also consider the probability 
$w_\ell^{(n)}(m)$ that the two marked edges at distance $\ell$
be separated by exactly $m$ minimal necks. This probability is
computed in Appendix A in the limit $n\to\infty$. It
takes the simple form:
\eqn\simpleform{w_\ell^{(\infty)}(m) \to {16\over 81} (m+1) \left({5\over 9}
\right)^m\quad {\rm for}\  \ell\to \infty\ .}
Note that $\sum\limits_{m\geq 0}{16\over 81} (m+1)\left({5\over 9}\right)^m=1$,
which means that, even if the two marked edges are far apart, there
is still a finite number of minimal necks to go through to reach 
one from the other. Now for $\ell\to\infty$, the two marked
edges lie with probability $1$ into two different baby universes and 
we moreover expect that a geodesic
path between them goes through, say $m_1$ minimal necks
in the vicinity (i.e.\ at a distance which remains finite when 
$\ell\to \infty$) of the first edge, then  
travels a distance of order $\ell$ in the mother universe, and
finally goes through $m_2$ minimal necks in the vicinity 
of the second edge. The distribution of minimal necks 
in the vicinity of one of the two edges is a local property which 
does not depend on the second edge (provided it is far enough) and simply
describes how the first edge is linked to the mother universe. 
The law $\wp(m_1)$ for $m_1$ and that for $m_2$ are the same by symmetry, 
and we have
\eqn\wwp{{16\over 81} (m+1)\left({5\over 9}\right)^m =\sum_{m_1+m_2=m}
\wp(m_1)\, \wp(m_2)}
from which we deduce the law for the number $m_1$ of minimal necks to 
go through to reach the mother universe from the first edge: 
\eqn\wplaw{\wp(m_1)={4\over 9}\left({5\over 9}\right)^{m_1}\ .}
In particular, we have a probability $\wp(0)=4/9$ that the first component of
the kernel, i.e.\ that closest to the first edge be the mother universe itself .
Note that this probability is slightly bigger than the known probability $1/3$
that the a marked edge belongs to the mother universe. This is because, 
with our definition of kernel, it is sufficient for having $m_1=0$ that 
the first edge has the same extremities as an edge belonging to the mother 
universe. Note that the law $\wp(m_1)$ is in practice a property
of simply rooted quadrangulations and that the introduction
of the second edge at distance $\ell$ was only instrumental in
the calculation.

In the same spirit, we may more precisely study the law for the
{\it distance from the origin of the
root edge to the mother universe} in large rooted quadrangulations. 
The detailed calculation is presentation in Appendix A. When $n\to \infty$,
we find a probability
\eqn\piofD{\pi(D)={4(5+2D)\over (D+2)^2(D+3)^2}}
that the root edge be at a distance $D$ from the mother universe.

\newsec{Discussion and conclusion}

In this paper, we have presented a detailed calculation of the two-point
function for quadrangulations with no multiple edges, based on a coding
of these maps by well-balanced well-labeled trees. These trees could be
enumerated exactly, for instance via a simple substitution procedure 
relating the known 
generating functions ($Q_\ell$, $F_\ell$ and $H_\ell$) for regular 
well-labeled trees to generating functions ($q_\ell$, $f_\ell$ and $h_\ell$) 
for almost well-balanced trees and by then extracting from these generating
functions the contributions ($p_\ell$, $e_\ell$ and $g_\ell$) of fully 
well-balanced ones. 

In the scaling limit, the two-point function of
quadrangulations with no multiple edges of size $n$ matches precisely that 
of general quadrangulations of size $3n$. This is consistent with the
picture of a general quadrangulation of size $3n$ as made of a mother universe 
with no multiple edges of size $n$ with attached baby-universes of 
size negligible with respect to $n$. In large general quadrangulations, 
two generic points at a large mutual distance will lie in different baby 
universes so that the geodesics between them will travel mostly within 
the mother universe. More precisely, the probability that two marked points 
at distance $\ell$ lie in the same baby universe tends to zero 
like $1/\ell^2$ and by taking $\ell\to \infty$, we may therefore guarantee 
that any path linking them enters the mother universe. We used this property 
to compute the law for the distance of a random point to the mother universe 
or for the number of necks to go through to reach this mother universe.

A different two-point function is obtained by conditioning the two 
marked points to lie in the same baby universe, which can be done by
requiring that the associated kernel be finite. In this case, the geodesics 
between the two points do not enter the mother universe and this leads to  
the a new two-point function {\it inside} a baby universe, which we computed 
in the local limit. At large distance $\ell$, it growths as $\ell$ instead 
of $\ell^3$. 

Our results deal with minimal neck baby universes only, which, 
as apparent on the two-point function, disappear in the scaling limit of 
large maps. Still, a more general baby universe structure should remain
visible in the scaling regime, involving larger necks of size of
order $n^{1/4}$. This structure is revealed for instance by the 
phenomenon of confluence of geodesics, but no precise rigorous statement was 
made so far to corroborate this picture.  In this respect, it would of
course be desirable to be able to eliminate cycles of arbitrary size in large 
quadrangulations. We however here face the problem of having a canonical 
prescription of which cycles to eliminate and which to retain in the mother 
universe. The case of cycles of length $4$ in quadrangulations with no multiple
edges is still tractable and removing non-contractible cycles of length
$4$ is known to correspond in the equivalence with general maps to going from 
$2$-connected to $3$-connected ones. Keeping track of distances while
removing these cycles seems to be a tractable issue. 

A much simpler question is that of more general
maps, for instance $2p-$angulations ($p>2$) with no multiple edges.
We have indeed a well-labeled mobile description of these maps \MOB\ in
general and removing double edges again simply amounts to making
these mobiles well-balanced. By a simple substitution procedure, 
we can obtain the generating function $p_1$ of rooted $2p$-angulations 
with no multiple edges with a weight $z$ per face:
\eqn\ponetwop{\eqalign{p_1&=z {2p-1\choose p} r^{p-1}-{2p-1\choose p+1} r^p
\cr {\rm with}&\quad r=1+z {2p-1 \choose p+1} r^{p+1}\ .\cr}}
and there seems to be no technical problem to address the question of
the distance-dependent two-point function in this case.

Finally, it seems also possible to address the question of the three-point 
function in quadrangulations with no multiple edges. Again we expect to
recover at large distances that of general quadrangulations, up to 
a global rescaling of the size by a factor of $3$.

\appendix{A}{}
We consider here the ensemble of quadrangulations with
two marked edges at distance $\ell$ and with size $n$. The
probability $w_\ell^{(n)}(m)$ that the two marked edges 
be separated by exactly $m$ minimal necks reads (for $\ell\geq 1$):
\eqn\probnumnec{w_\ell^{(n)}(m)= {
\left.R_1^2\,\left(\delta_{m,0}\!+\!\left[{\hat g}\left(t,g\, R_1^2\right)\right]^{m+1}\right)
\right\vert_{t^{\ell-1}\, g^n}
\over H_\ell\vert_{g^n}} = {\left. R_1^2\, \left(\delta_{m,0}\!+\!\left[1\!-\!{R_1^2\over {\hat H}(t,g)}
\right]^{m+1}\right)\right\vert_{t^{\ell-1}\, g^n}
\over H_\ell\vert_{g^n} }}
where we used ${\hat g}(t,z(g))=1-1/{\hat h}(t,z(g))=1-R_1^2/{\hat H}(t,g)$
upon introducing  
\eqn\Hhat{{\hat H}(t,g)\equiv \sum_{\ell\geq 0} H_{\ell+1}(g)\, t^{\ell}\ .}
To extract the large $n$ leading behavior of the $g^n$ coefficients, 
we use the expansions
\eqn\expanrone{R_1^2= {16\over 9}(1-2\epsilon^2+4 \epsilon^3 +\cdots)}
and 
\eqn\expanHhat{{\hat H}(t,g)={\hat A}^{(H)}(t)+{\hat C}^{(H)}(t) \epsilon^2 
+{\hat D}^{(H)}(t)\epsilon^3+\cdots}
with, in particular  
\eqn\limtone{\eqalign{{\hat A}^{(H)}(t)&=
{4 \left(-3 t^2+4 {\rm Li}_2(t) t+4 t-4 {\rm Li}_2(t)\right)\over t^3}
\buildrel {t\to 1}\over \rightarrow
4 \cr {\hat D}^{(H)}(t) & = 
\sum_{\ell\geq 0}D_{\ell+1}\, t^\ell 
\buildrel {t\to 1} \over \sim {96\over 7 (1-t)^4}
\ .\cr}}
Picking the leading singularity (coefficient $\propto \epsilon^3$) of both
the numerator and the denominator of \probnumnec, we deduce that, 
for $n\to \infty$,
\eqn\probnumnecasy{\eqalign{
w_\ell^{(\infty)}(m)& =
{\left.{16\over 9}\left[1\!-\!{16\over 9 {\hat A}^{(H)}(t)}\right]^m
\left[4\left(1\!-\!{16\over 9 {\hat A}^{(H)}(t)}\!+\!\delta_{m,0}\right)\!
+\!{16 (m+1)\over 9 
{\hat A}^{(H)}(t)}
\left(
{{\hat D}^{(H)}(t)\over {\hat A}^{(H)}(t)}\!-\!4\right)
 \right]\right\vert_{t^{\ell-1}}
\over {\hat D}^{(H)}(t)\vert_{t^{\ell-1}}}
\cr & \buildrel {\ell \gg 1} \over 
\sim \left({16\over 9 {\hat A}^{(H)}(1)}\right)^2 (m+1) 
\left[1\!-\!{16\over 9 {\hat A}^{(H)}(1)}\right]^m
={16\over 81} (m+1)\left({5\over 9}\right)^m
\cr}}
since the large $\ell$ behavior is entirely dictated by the $t\to 1$ limit.

We may alternatively study the law for the
{\it distance from the origin of
root edge to the mother universe} in large rooted quadrangulations. 
It may be obtained by first computing the law for the distance to
the first minimal neck to go through 
to reach the mother universe. We call this distance $d$ if the 
first minimal neck is formed of two $d\to (d+1)$ edges.  We have
in particular $d=0$ if the first neck consists of two $0\to 1$ edges, 
with the extremity at distance $1$ necessarily different from the extremity 
of the root edge.
By convention, if the root edge already ``touches'' the mother universe
(i.e.\ belongs to the mother universe or has the same extremities as an edge 
belonging to the mother universe) we set $d=\infty$ as there is no
first minimal neck in this case.
The probability law for $d$ may again be obtained by temporarily introducing 
a second marked edge at distance $\ell$ and then sending $\ell\to\infty$.
It reads:
\eqn\probd{\eqalign{P(d)& = \lim_{\ell\to \infty} 
\lim_{n\to \infty} {\left. R_1^2\ g_{d+1}\left(g\, R_1^2\right)\, h_{\ell-d}
\left(g\, R_1^2\right)\right\vert_{g^n}\over
H_\ell\vert_{g^n}}\cr
&= 
\lim_{\ell\to\infty}
{\left.\left(1\!-\!{16\over 9 {\hat A}^{(H)}(t)}\right)\right\vert_{t^d}
{\hat D}^{(H)}(t)\vert_{t^{\ell-d-1}}\!+\!
\left.\left({16\over 9 {\hat A}^{(H)}(t)}
\left(
{{\hat D}^{(H)}(t)\over {\hat A}^{(H)}(t)}\!-\!4\right)
 \right)\right\vert_{t^d}{\hat A}^{(H)}(t)\vert_{t^{\ell-d-1}}
\over {\hat D}^{(H)}(t)\vert_{t^{\ell-1}}}
\cr
&= \left.\left(1\!-\!{16\over 9 {\hat A}^{(H)}(t)}\right)\right\vert_{t^d}\cr}
}
or equivalently 
\eqn\equiprobd{\sum_{d\geq 0}P(d)\, t^d=1-{16\over 9 {\hat A}^{(H)}(t)}
= {1\over 5}+ {7\over 25}t + {79\over 2500} t^2 +{699\over 50000}t^3 +
{1910211\over 245000000} t^4 +\cdots
}
where we explicited the first values of $P(d)$ for $d=0,1,2,3,4$. 
We have in particular a probability 
$\sum\limits_{d\geq 0} P(d)=5/9$ that we have to go through at least a minimal 
neck to reach the mother universe and, if so, the distance
to this minimal neck is on average 
$\sum\limits_{d\geq 0} d\, P(d)/\sum\limits_{d\geq 0} 
P(d)= (4/5) (2 \pi^2 /3-5)= 1.26379...$. The complementary probability $4/9$
is that of the event of a root edge touching the mother universe. 
This is consistent with the value $\wp(0)=4/9$. For large $d$, we find from 
the singular behavior of $A^{(H)}(t)$ at $t=1$ that $P(d)/\sum\limits_{d\geq 0}
P(d)\sim 32/(5 d^3)$. Note that along the same lines, we can as well
compute the joint probability
for the distance $d$ to the first minimal neck and for the size $k$ 
of the component linking the root edge to this neck. We find a probability
$g_{d+1}\vert_{z^k} (4/27)^k$, which, as it should adds up to $P(d)$ upon 
summing over $k$. 

Consider now the total distance $D$ from the origin of the root edge to the
mother universe in large quadrangulations. This distance is $0$ if
the root edge touches the mother universe or if it is separated from
the mother universe only by (nested) necks made of pairs of $0\to 1$ edges
only. The law for $D$ may be deduced from $P(d)$ above upon writing
\eqn\pilaw{\pi(D)={4\over 9}\delta_{D,0}+P(D) {4\over 9}+
P*P(D) {4\over 9}+ P*P*P(D) {4\over 9}+\cdots }
where $P*Q(D)=\sum_{d=0}^D P(D)Q(d-D)$ is a simple convolution. 
Indeed, any geodesic path to the mother universe generically crosses a number
$m$ of necks and may accordingly be decomposed into $m$ pieces whose
lengths add up to $D$ and share {\it the same probability law} $P$.  
This leads to a probability $P^{*m}(D)$, to be summed over $m$. 
We get finally
\eqn\probD{\eqalign{\sum_{D\geq 0}\pi(D)\, t^D & = 
{4\over 9}{1\over 1-\sum_{D\geq 0}P(D)}\, t^D\cr 
&
={{\hat A}^{(H)}(t)\over 4}
=\sum_{D\geq 0} {A_{D+1}\over 4}\, t^D \cr
}}
with in particular $\sum\limits_{D\geq 0}\pi(D)=1$. We deduce
\eqn\valpiD{\pi(D)={A_{D+1}\over 4}= {4 (5+2D)\over (D+2)^2(D+3)^2}\ .}
In particular, the average distance is $2 \pi^2/3-5=1.57974...$
and for large $D$, we have $\pi(D)\sim 8/D^3$.

\listrefs
\end